\documentclass[useAMS,usenatbib]{mn2e}

\usepackage{graphicx}
\usepackage{euscript}
\usepackage{amsmath}
\usepackage{amsfonts}
\usepackage{multicol}

\usepackage{multicol}
\usepackage{multirow}
\usepackage{subfigure}

\newcommand{\intl}{\int\limits}

\title[Funnel radiation in SS 433]
{Evidences of the supercritical disc funnel radiation in X-ray spectra of SS~433}
\author[A. Medvedev, S. Fabrika]
  {A.~Medvedev,$^1$ S.~Fabrika$^2$ \\
  $^1$ Moscow State University, Russia, Moscow\\
  $^2$ Special Astrophysical Observatory, Russia, Nizhij Arkhyz}
\date{Accepted 2009 October 16. Received 2009 September 23; in original form 2009 June 03}

\pagerange{\pageref{firstpage}--\pageref{lastpage}} \pubyear{2009}

\def\LaTeX{L\kern-.36em\raise.3ex\hbox{a}\kern-.15em
    T\kern-.1667em\lower.7ex\hbox{E}\kern-.125emX}

\begin{document}

\label{firstpage}

\maketitle

\begin{abstract}
We have analysed the XMM-Newton spectra of SS\,433 using a standard model of 
adiabatically and radiatively cooling X-ray jets. The multi-temperature thermal
jet model reproduces well the strongest observed emission line fluxes. Fitting the He-
and H-like iron line fluxes, we find that the visible blue jet base temperature
is $\approx 17$~keV, the jet kinetic luminosity $L_k \sim 2 \cdot 10^{39}$\,erg/s
and the absorbing column density $N_H \sim 1.5 \cdot 10^{22}$\,cm$^{-2}$.
All these parameters are in line with the previous studies. The thermal model
alone can not reproduce the continuum radiation in the XMM spectral range, the
fluorescent iron line and some broad spectral features. Using the thermal 
jet-plus-reflection model, we find a notable
contribution of ionized reflection to the spectrum in the energy range
from $\sim 3$ to 12~keV. The reflecting surface is highly ionized
($\xi \sim 300$), the illuminating radiation photon index changes from
the flat spectrum ($\Gamma \approx 2$) in the 7\,--\,12~keV range to
$\Gamma \approx 1.6$ in the range of 4\,--\,7~keV, and to $\Gamma \la 1$ in
the range of 2\,--\,4~keV. We conclude that the reflected spectrum
is an evidence of the supercritical disc funnel, where the illuminating
radiation comes from deeper funnel regions, to be further reflected in the outer
visible funnel walls ($r\ga 2 \cdot 10^{11}$ cm). In the multiple scatterings
in the funnel, the harder radiation $> 7$~keV may survive absorption,
but softer radiation is absorbed, making the illuminating spectrum curved.
We have not found any evidences of reflection in the soft 0.8\,--\,2~keV
energy range, instead, a soft excess is detected, that does not depend
on the thermal jet model details. However the soft component spectrum
is basically unknown. This soft component might prove to be the 
direct radiation of the visible funnel wall. It is represented here either 
as black body radiation with the temperature of $\theta_{bb} \approx 0.1$~keV 
and luminosity of $L_{bb} \sim 3 \cdot 10^{37}$\,erg/s, or with a multicolour 
funnel (MCF) model. The soft spectral component has about the same parameters 
as those found in ULXs.

\end{abstract}


%%%%%%%%%%%%%%%%%%%%%%%%%%%%%%%%%%%%%%%%%%%%%%%%%%%%%%%%%%%%%%%%%%%%%%%%%%%%%%%%%%%
\section{Introduction}
\label{S:Intro}
%%%%%%%%%%%%%%%%%%%%%%%%%%%%%%%%%%%%%%%%%%%%%%%%%%%%%%%%%%%%%%%%%%%%%%%%%%%%%%%%%%%

SS\,433 is the only known persistent superaccretor in the Galaxy -- a source of 
relativistic jets \citep[][for review]{Fab04}. This is a massive close binary, 
where the compact star is most probably a black hole
\citep{Giesetal02,Cheretal05,Hill_Gies08,Blundelletal08}.
SS\,433 intrinsic luminosity is estimated to be $\sim 10^{40}$\,erg/s, with
its maximum located in non-observed UV region. In effect this is a very blue, 
but heavily absorbed object; its estimated temperature depends on the accretion 
disc orientation \citep{Murdin_Clark80,Cheretal82,Dolanetal97}, being 
$\sim 50000 - 70000$\,K when the disc is the most open to the observer. 
It is important that almost all the observed radiation
is formed in the supercritical accretion disc, and the donor star contributes 
less than 20\,\% of the optical radiation. The system's extreme luminosity 
suggests that the black hole's mass is $\sim 10$\,M$_{\sun}$.
Such a big energy budget of the object is supported by a very well 
measured kinetic luminosity of the jets, $\sim 10^{39}$\,erg/s, both in
direct X-ray and optical studies of the jets and
in the studies of the jet-powered nebula W\,50 \citep{Fab04}. At the
same time we know that practically all the energy at the accretion onto 
a relativistic star is released in X-rays. This means that the observed 
radiation of SS\,433 was thermalised in the strong wind coming from the 
supercritical disc.

The wind from a supercritical disc was first predicted by \citet{ShakSun73}
and later confirmed in radiation-hydrodynamic simulations
\citep{Eggumetal88,Okudaetal05,Ohsuga05}.
These ideas led to a prediction that SS\,433, being observed face-on appears as 
an extremely bright X-ray source, and we may expect an appearance of a new type of X-ray
sources in galaxies \citep{Katz87,FabMesch00,FabMesch01} -- the face-on SS\,433 stars.
It is quite possible that the new type of X-ray sources, ULXs (ultraluminous X-ray sources,
\citet{Roberts07}) are SS\,433-like objects observed nearly face-on. Their observed X-ray
luminosities are $10^{39} - 3 \cdot 10^{41}$\,erg/s and they are certainly related to
the massive star population. The disc orientation in SS\,433 is about edge-on, the 
angle between the disc plane and the line of sight is $10 \pm 20^{\circ}$, due to the
precessional variations. Therefore we have no chance to observe the funnel in the
supercritical disc directly.

The supercritical disc radiation is not isotropic. If one takes into account 
the geometrical beaming in the disc funnel with the (half) opening angle 
$\vartheta_f \sim 25^{\circ} - 30^{\circ}$ \citep{Ohsuga05} 
and that a supercritical disc surface radiation is local Eddington \citep{ShakSun73}, 
one finds \citep{Fab_etal06}, that the observed X-ray luminosity of 
$\sim 10^{41}$\,erg/s is expected for the face-on SS433 star. 
The recent data show \citep{Stobbart06,Berghea08} that ULXs posses curvature and rather 
flat X-ray spectra, which are difficult to interpret with a single-component or
any other simple model. The supercritical accretion discs are expected to have flat 
($\nu F_{\nu} \propto \nu^0$) X-ray spectrum \citep{Poutan07}, because the energy release
$Q(r)$ in the discs must be $Q(r) \propto r^{-2}$ ($T(r) \propto r^{-1/2}$) to make 
the disc thick due to the radiation pressure.
%and to produce strong wind. 

The X-ray luminosity of SS\,433 is $\sim 10^{36}$\,erg/s (at a distance of 5.5\,kpc 
\citep{Blund_Bowler04}), four orders of magnitude less than the bolometric luminosity.
It is believed that all the X-rays come from the cooling X-ray jets. The jet may be easily
accelerated by the radiation in the hydrodynamic funnel \citep[e.g., ][]{Ohsuga05} of the 
supercritical disc. The jet velocity value, $v_j \approx 0.26 c$, and its unique stability,
where the velocity does not depend on the activity state, indicate that the jet 
acceleration must be controlled by the line-locking mechanism \citep{shapiro86,Fab04}. 
The funnel has to be relatively transparent for the accelerating
radiation, as it was confirmed in radiation-hydrodynamic simulatuons
\citep[e.g., ][]{Ohsuga05}. 

These reasonings, however, bear a problem, why do not we observe any
funnel radiation, missing four orders of magnitude in the X-rays? Even a small amount of 
gas in the most outer part of the funnel, where we observe the X-ray jet base, may reflect 
and scatter some part of the direct $\sim 10^{40}$\,erg/s of the funnel radiation. The structure 
of the funnel and the jet acceleration/collimation region is unknown. Apparently,  
the direct funnel radiation is entirely blocked for the observer. The 'standard' jet 
model \citep{brinkmann88,kotani96,marshall02,filippova06} suggests that all the 
observed X-ray radiation of SS\,433 is formed in the cooling X-ray jets. However, an 
analysis of the latest XMM observations \citep{brinkmann05} led the authors to a conclusion
that the 'standard' jet model can not produce the observed X-ray continuum. 

In the 'standard' jet model there are two conical anti-parallel jets, considered identical.
The jet's gas is optically thin, being in collisional ionization equilibrium 
and cooling adiabatically (or adiabatically plus radiatively). The jets are observed beginning 
from the jet base $r_0$ (the distance between the
base of the visible jet and the black hole, different for the red and blue jets), 
which depends on the precessional phase and on 
the eclipses by the donor star. The gas temperature at the jet base $T_0(r_0)$ is estimated from 
observations. Precessional and orbital variability of SS\,433 is well-known \citep{Fab04},
the jets (and both the accretion disc, and the accretion disc wind) precess with a 164-day period
and an amplitude of $\pm 20^{\circ}$. At the precessional phase $\psi \approx 0$ the disc is
the most open to the observer, the angle between the jets and line of sight is $\sim 60^{\circ}$
(we refer here to the approximate values because there are nutation-like and sporadic variabilities,
both of about $3-5^{\circ}$). The accretion disc rim (or the opaque  part of the wind) is thick
$h/r \sim1$ \citep{filippova06}. Comparing the amplitudes of the precessional and orbital
variabilities in different X-ray bands, \citet{Cheretal05} found that both the donor star and
the outer disc rim have about the same size, Therefore, one may expect the same radius for 
the approaching jet base $r_0 \sim 10^{12}$\,cm.

In X-ray observations with ASCA and Chandra \citep{kotani96,marshall02}, dozens of X-ray
emission lines formed both in blue and red jets were resolved. The lines are variable both 
in positions (in accordance with the jet kinematic model) and in intensities, the strongest are
He- and H-like $\text{Fe XXV\,K}\alpha$ and $\text{Fe\,XXVI\,L}\alpha$ iron lines. This 
allowed the 'iron line diagnostics' of the gas temperature at the jet bases by measuring 
the line ratio Fe\,XXV/Fe\,XXVI. The temperature at the base of the jet was estimated as 
$\theta_0 = kT_0 = 10 \div20$~keV \citep{kotani96,marshall02,namiki03}. In the latest 
XMM-Newton observations \citet{brinkmann05} estimate the temperature of about 
$\theta_0 \sim 17 \pm 2$~keV, with the main uncertainty coming from the form of the
underlying continuum. Fitting the continuum with a thermal bremsstrahlung model 
\citep[GINGA data, ][]{brinknann91} gives a notably bigger value, $\theta_0 \ga 30 $~keV.
The jet base temperature determined both in the line diagnostics and in the continuum fits,
drops notably in eclipses, what confirms the cooling jet model. Analysing the numerous
ASCA data, \citet{kotani96} found that the farther part of the receding jet is weakened
(probably absorbed in the external gas located in the disc plane) and Nickel is highly 
overabundant ($\sim 10$ times) in the jets. The last finding has been confirmed 
in the XMM observations \citep{brinkmann05}. Note that some discrepancies may exist 
as the system SS\,433 is highly variable and its appearance depends strongly on
the precessional phase.

\citet{brinkmann05} found that the XMM SS\,433 spectrum can not be fitted with 
any simple continuum law. The numerous lines point at the thermal model, but a 
strong continuum curvature indicates the Comptonization. The best formal 
model of the additional continuum component is a broken power law with 
a break at $\sim 7.1$~keV, where the low energy PL rises with energy 
($\Gamma \sim -1$), the high energy part is rather steep ($\Gamma \sim 4$), 
however, the parameters are not usually well determined in the fits. The overall 
continuum is too hard to be bremsstrahlung, the electron-electron bremsstrahlung
can not fit it even for the temperatures of 40~keV. A highly absorbed additional thermal 
component gives a too strong $\sim 7$~kev edge. \citet{brinkmann05} discuss a model
with a single, very broad $\sim 7$~kev line, which may be formed in the innermost hot jet
(otherwise hidden) and Compton down-scattered by colder material. These uncertainties
in continuum make problems for the iron-line diagnostics of the jet base temperature.

Practically in all the studies of SS\,433 X-ray spectra the jet mass loss $\dot M_j$
and the jet kinetic luminosity $L_k = \dot M_j v_j^2 /2$ were determined. To estimate
$L_k$ one needs to adopt the jet (half) opening angle $\vartheta_j$. The mass loss 
rate derived from the X-ray line intensities depends strongly on the opening angle 
adopted, i.~e. gas density at the jet base, because the emission measure is 
$\propto n^2$. The kinetic luminosity values derived from X-ray spectra were quite 
diverse, up to $10^{42}$\,erg/s. However, in two latest studies, one from the Chandra 
data \citep{marshall02} and the other from the XMM data \citep{brinkmann05}, the values 
are $\sim 3 \cdot 10^{38}$ and $\sim 5 \cdot 10^{39}$\,erg/s respectively. They 
are even closer to one another, matching different distances to SS\,433 adopted 
by these authors. \citet{panferov} found $L_k \sim 10^{39}$\,erg/s from the 
jet Balmer line intensities in optical spectra.
The mass loss rate in the jets was estimated \citep{zealey,FabrBor87,dubner}
on the base of W\,50 nebula powered by the jets with approximately the same result,
$L_k \sim 10^{39}$\,erg/s. Thus the jet kinetic luminosity was measured with 
relatively good accuracy. 

The jet opening angle was directly measured in the studies of X-ray and optical line 
widths \citep{Fab04}. \citet{marshall02} found $\vartheta_j = 0.61 \pm 0.03^{\circ}$
from the Chandra spectra. They assumed that the density is uniform through the jet 
cone's cross section and $\vartheta_j$ is the jet's cross sectional radius. 
\citet{borfab87} found $\vartheta_j = 0.82 \pm 0.14^{\circ}$ from 
optical spectra, assuming that the emitting gas density is Gaussian-distributed 
in the cone's cross section. We give here the values of $\vartheta_j$ 
recalculated, using the \citet{marshall02} definition. Such a coincidence  
in the X-ray and optical jet opening angles makes the jets pure ballistic 
from X-ray ($\sim 10^{12}$\,cm) to optical ($\sim 10^{15}$\,cm) regions. 
In optical spectra the jet opening angle 
has been estimated from H$\alpha$ line profiles during 8-day long observations,
the nutation broading has been taken into account and possible jet sporadic 
activity was diminished. In other Chandra observations \citep{namiki03,lopez06} 
the opening angle has been found twice as big and it was greater in the higher 
temperature (Fe), than in the lower temperature (Si) parts of the jets. Regarding
comparatively short time scales of the X-ray observations and probable line broading
due to electron scattering \citep{namiki03} in the hottest parts of the jet,
we may adopt $\vartheta_j \approx 0.7^{\circ}$ as a good estimate of the jet opening 
angle.

In this paper we analyse X-ray spectra of SS\,433 using XMM data and the 'standard'
jet model. We do not try to determine the jet kinetic luminosity, because the 
X-ray continuum is complex and can not be explained in the 'standard' jet model 
\citep{brinkmann05}. Instead we adopt these two main parameters -- $L_k = 10^{39}$\,erg/s
($\dot M_j = 3.3 \cdot 10^{19}$\,g/s) and $2\vartheta_j = 1.5^{\circ}$, as 
well established and known. We find that the line fluxes can be well matched with 
the 'standard' jet model and study the additional components of the X-ray spectrum
of SS\,433.

%%%%%%%%%%%%%%%%%%%%%%%%%%%%%%%%%%%%%%%%%%%%%%%%%%%%%%%%%%%%%%%%%%%%%%%%%%%%%
\section{Observations and data reduction}
\label{S:Observ}
%%%%%%%%%%%%%%%%%%%%%%%%%%%%%%%%%%%%%%%%%%%%%%%%%%%%%%%%%%%%%%%%%%%%%%%%%%%%%

We used a public archive of the XMM-Newton 
Observatory.\footnote[1]{\texttt{http://xmm.vilspa.esa.es/external/xmm\_data\_acc/xsa/}}
There are 8 observations of SS\,433 in the archive, where the target was in the 
detector's centre. They were carried out in April 2003 (4 observations at the precessional
phase $\psi \approx 0.8$), October 2003 (2 observations at $\psi \approx 0$),
and two single observations in March 2004 and April 2004.
%($\psi \approx 0$) %($\psi \approx 0.15$). 
In this paper we concentrate on the study of X-ray jets and the thick accretion disc 
(its wind). Then we have selected only those observations, where 
disc was not eclipsed by the donor and was the most open to the observer.
This is the observation taken on 19 October 2003 in orbit 707 ($\psi = 0.04$) 
in the orbital phase $\varphi = 0.28$ (for orbital and precessional phases 
see \citet{Fab04}). For comparison purposes we selected the observation from
6 April 2003 in orbit 609, taken at the intermediate inclination ($\psi \approx 0.84$)
of the disc and at about the same orbital phase $\varphi = 0.34$. The last observation
is relatively short, hence we added the following observation taken  
in the same accretion disc precessional position, in orbit 610 ($\varphi = 0.49$)
for a control check. 

These data allow us to study the jets and the disc wind at two different 
orientations of the disc/jets and at about the same orbital phase. With such a strong 
mass transfer in the system, the jet and the accretion disc funnel visibility 
behind the wind photosphere (the outer disc rim) may be dependent on the orbital 
phase. 

We use the EPIC PN observations to get the highest signal-to-noise data. During all the
observations the EPIC PN camera was operated in the Small Window mode with a medium 
filter. The PN data were reprocessed using the XMMSAS version 6.1. The actual Life
Time of the detector after a correction for Good Time Intervals was 11.4\,ksec in orbit
707, 4.4\,ksec in orbit 609 and 5.6\,ksec in orbit 610. 

%%razmer diafragmy extractsii?

%%%%%%%%%%%%%%%%%%%%%%%%%%%%%%%%%%%%%%%%%%%%%%%%%%%%%%%%%%%%%%%%%%%%%%%%
\section{Jet model}
\label{S:Model}
%%%%%%%%%%%%%%%%%%%%%%%%%%%%%%%%%%%%%%%%%%%%%%%%%%%%%%%%%%%%%%%%%%%%%%%%

To model the jet spectrum we use the standard approach which was used by many 
authors to analyse the X-ray spectra of SS\,433 
\citep{kotani96,brinknann91,brinkmann05,filippova06}. 
This standard multi-temperature model describes well both the jet line spectrum and 
its variability. The jet gas moves in ballistic trajectories with a velocity 
$v = 0.26c$. In such a case the gas concentration can be written as follows:
\begin{equation}\label{E:n2}
n(r) = n_0\left(\frac{r}{r_0}\right)^{-2}, 
\quad\text{where}\quad n_0 = \frac1{\Omega_j}\frac{\dot M_j}{\mu m_p  v_j r_0^2}.
\end{equation}
Here $r_0$ is the visible jet base (the distance between the
base and the black hole), $\Omega_j = 2\pi(1 - \cos \theta_j)$ is the solid opening angle of the jet,
$\dot M_j$ -- the mass loss rate in the jet, $m_p$ -- the mass of a proton, $\mu$ -- molecular weight.
The gas temperature $\theta = kT$ along the jet accounting for adiabatic and radiation cooling 
is determinated by the equation: 
\begin{equation}\label{E:thermo2}
\frac32\frac{d\theta}{\theta} - \frac{dn}{n}\left(1 + 
\frac{r \widetilde\varepsilon}{2v_j\theta}\right) = 0,
\end{equation}
where $\widetilde\varepsilon$ stands for specific energy losses by radiation per one particle. 
Introducing new variables $\tau = \theta/\theta_0$, $x=r/r_0$, $\widetilde\varepsilon = \mu nJ$
one may write the equation (\ref{E:thermo2}) in a dimensionless form:
\begin{equation}\label{E:Teq}
 \frac{d\tau}{dx} = -\frac43\frac{\tau}{x} - \alpha\frac{J(\tau; \theta_0)}{x^2},
\end{equation}
where $\alpha$ is a coefficient depending on physical parameters of the jet $\dot M_j$ and $r_0$:
\begin{equation}
 \alpha = \frac{\mu}{\Omega_j m_p v_j^2}\frac{\dot M_j}{r_0\theta_0}.
\end{equation}
%�� alpha ������ mu, ��������� eps = n_H J = mu n J
%%%%%%%%% net, ne ubrano! prover'  %%%%%%%%%%%%%%

The introduced value $J(\theta)$ is the total emissivity of the plasma, and radiation losses 
integrated over all energies,
\begin{equation}
 J(\theta) = \int J_\epsilon (\theta) d\epsilon.
\end{equation}
The value $J(\theta)$ is tabulated in the model of hot, optically thin plasma APEC/APED, 
which uses the atomic database ATOMDB v1.3.1 (\citealt{smith1}, \citealt{smith2}, \citealt{smith3}).
ATOMDB provides an improved spectral modelling capability through additional emission lines, 
accurate wavelengths for strong X-ray transitions, and new density-dependent calculations 
(for more details see http://cxc.harvard.edu/atomdb/). Relativistic boosting is taken
into account in this model. The model provides a possibility of separate analysis of both 
line and continuum emissivities.

To calculate the multi-temperature jet spectrum we divide the jet in 20 equal parts in 
the section $r_0 - 4r_0$, the gas temperature at $\sim 4r_0$ drops below 100\,eV, 
where the thermal instabilities begin to operate \citep{kotani96}. One needs to consider 
many (at least 20) mono-temperatute parts of the jet, because the radiative cooling 
begins to dominate the adiabatic cooling at temperatures below $\sim 2$\,kev and 
has a strong peak at $\sim 0.6$\,kev. The total spectrum of the jet is calculated 
using the formula:
\begin{equation}\label{E:Jetspecta}
F_\epsilon =  \sum_i N_i J_{\epsilon}(\theta_i^*),
\end{equation}
where $J_\epsilon$ is the plasma emissivity and $N_i$ is the spectrum normalisation 
of $i$-th part of the jet:
\begin{equation}\label{E:Norm}
 N_i = \frac{1}{4\pi D^2}\intl_{V_i} n_e n_H dV,
\end{equation}
where $D$ is a distance to SS\,433 \citep[5.5\,kpc, ][]{Blund_Bowler04}.
%$n_e$ -- concentration of electrons, $n_H$ -- concentration of hydrogen atoms, 
The integration is carried out over the jet portion volumes $V_j$. The average gas temperature of the   
$i$-th part of the jet $\theta_i^*$ is estimated as follows:
\begin{equation}
 \theta_i^* = \frac{\intl_{r_i} \theta(r) J(r) \frac{dr}{r^2}}{\intl_{r_i} J(r) \frac{dr}{r^2}}.
\end{equation}

The main parameters which define the physical state of the jet are gas 
density and temperature $\theta_0(r_0)$ at the visible jet base. The density 
depends on the mass loss rate $\dot M_j$ and the jet (half) opening angle 
$\vartheta_j$. As we discussed in the Introduction, the last two
parameters were relatively well measured in SS\,433 in previous studies not only in 
X-ray range. The opening angle was particularly very well measured in optical studies.
To make our analysis more certain we do not try to determine the jet kinetic luminosity
(the mass loss rate) from the XMM spectra, the more so, that one has serious problems with 
an interpretation of continuum at $\sim 7$~keV \citep{brinkmann05}. Therefore, we adopt 
the parameters $L_{kin} = 10^{39}$\,erg/s and $2 \vartheta_j = 1.5^{\circ}$ as already 
known. 

From all previous X-ray studies we may expect the gas temperature at the base
is in the range of $\theta_0(r_0) = 8 - 20$~keV (see Introduction), and the visible jet base 
$r_0 \sim 10^{12}$\,cm. The last estimate is quite substantial, it was argued 
above, that both the donor star and the outer disc rim have about the same size
\citep{Cheretal05,filippova06}. The partial eclipses of the X-ray jets by the donor 
observed in SS\,433 prove the estimate of the X-ray jet base $r_0$.

In the thermal jet model we adopt the solar abundance of elements, however, Nickel is 
10 times overabundant \citep{brinkmann05}. Section \ref{reflection} discusses 
this in detail.  

\begin{figure} 
\begin{center}
 \includegraphics[width=80mm]{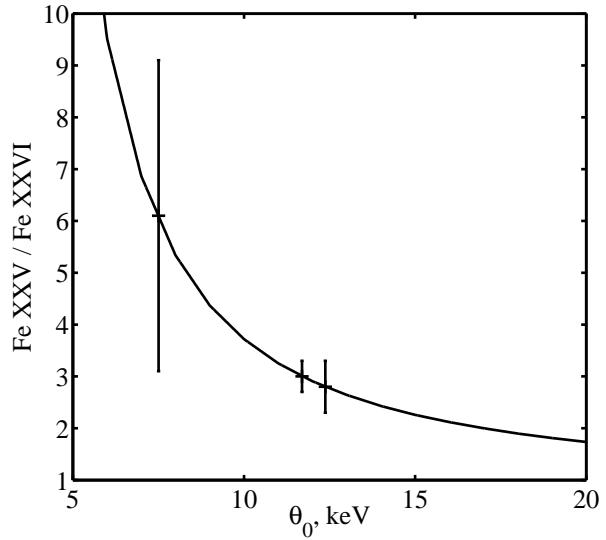}
\caption{Measured ratio of the blue jet Fe\,XXV and Fe\,XXVI line fluxes (Table\,1),
derived with the simple local continuum fit,
and the model ratio as depending on the gas temperature at the jet base. The highest 
temperature, $\theta_0 \sim 12$~keV, is obtained in orbit 707 ($\psi \approx 0.0$),
a slightly lower temperature in orbit 610 ($\psi \approx 0.8$); in orbit 609 we find
the temperature $\theta_0 \sim 8$~keV with notably bigger uncertainty.
}
\label{F:ratio}
\end{center}
\end{figure}

%%%%%%%%%%%%%%%%%%%%%%%%%%%%%%%%%%%%%%%%%%%%%%%%%%%%%%%%%%%%%%%%%%%%%%%%%%%%%%%%%%%
%%\section{Initial jet's parameters estimation via fluxes in the iron lines $\FeK$ and $\FeLy$}
\section{Jet parameters in iron line diagnostics}
\label{S:parameters}
%%%%%%%%%%%%%%%%%%%%%%%%%%%%%%%%%%%%%%%%%%%%%%%%%%%%%%%%%%%%%%%%%%%%%%%%%%%%%%%%%%%

In this section we estimate the jet parameters $\theta_0$ and $r_0$ using the fluxes 
and the flux ratio of the iron lines Fe\,XXV\,K$_\alpha$ and Fe\,XXVI\,Ly$_\alpha$ 
\citep{kotani96,brinkmann05}
%%%%%%%%%%%%%%%%%%%%%%%%%%%%VST
and adopting a simple local PL continuum. 
These Fe\,XXV and Fe\,XXVI lines formed in the approaching 
jet are the strongest and well resolved in the XMM spectra. The resulting line fluxes 
depend on the underlying continuum model. Therefore we use the simplest continuum to
analyse here how stable are the derived parameters depending on $\dot M_j$ and on 
the errors of the measured line fluxes. We do not examine here different continuum 
%%%%%%%%%%%%%%%%%%%%%%%%%%%%VST
models, because \citet{brinkmann05} have studied these XMM spectra using many 
versions of continuum and concluded that "the high sensitivity and wide bandpass of the 
XMM-Newton instruments rule out any of the simple continuum models used previously".
Instead in the section below we find clear evidences for a certain additional continuum 
component (reflection), study the whole spectrum with these two models, thermal jet 
and reflection, and correct the values of $\theta_0$ and $r_0$.

\begin{figure} 
\begin{center}
\includegraphics[width=70mm]{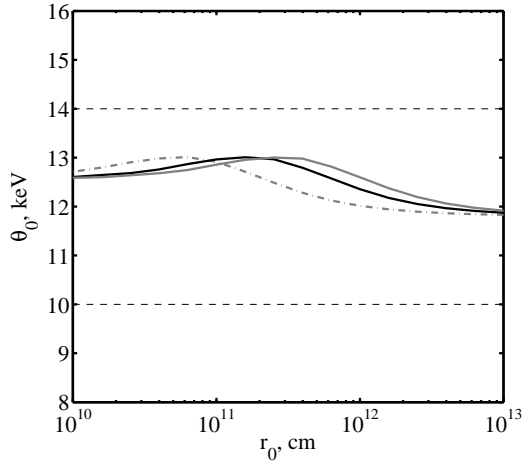}
\caption{Dependence of the derived gas temperature at the jet base on the variations of the 
visible base position $r_0$ 
%%%%%%%%%%%%%%%%%%%%%%%%%%%%VST
(in the simple local continuum model)
and mass loss rate $\dot M_j$ for orbit 707 observation ($\psi = 0$). 
The solid curve is for $\dot M_j = 3.3\times10^{19}$\,g/s ($L_{kin} = 10^{39}$\,erg/s), 
the dashed grey curve is for $1.5 \times 10^{19}$, the solid grey curve is for 
$5 \times 10^{19}$\,g/s. Two horizontal dashed lines mark the temperature limits 
(Table\,1). 
}\label{F:Tlimits}
\end{center}
\end{figure}

\begin{table*}
\caption{Observed fluxes and derived blue jet preliminary parameters in the simplest 
continuum model. The columns present precessional $\psi$ and orbital $\varphi$ phases 
of the observations, line fluxes in units $10^{-4}$\,ph/s\,cm$^2$, the flux ratios 
$\mathcal{R}$, the temperature at the jet visible base $\theta_0$ and the base $r_0$.
}
  \begin{tabular}{lllllll}
   \hline
$\psi$&$\varphi$&F(Fe\,XXV)&F(Fe\,XXVI)&$\mathcal{R}$&$\theta_0$,~keV&$r_0$, $10^{11}$\,cm\\
   \hline
0.04 & 0.28 & $5.5 \pm 0.4$ & $1.95\pm 0.3$& $2.8\pm0.5$ & $12\pm 2$      & $2.8 \pm 0.2  $  \\
0.84 & 0.34 & $4.9 \pm 0.7$ & $0.8 \pm 0.4$& $6.1\pm3$   & $8\pm^{4}_{1}$ & $2.6\pm 0.3$\\
0.84 & 0.49 & $5.9 \pm 0.2$ & $2.0 \pm 0.2$& $3.0\pm0.3$ & $12\pm 1 $     & $2.6\pm 0.1$\\
   \hline
  \end{tabular}
\end{table*}
\label{T:fluxes}

%%%%% Vse-taki troinye li oshibki uzakany, tam est nesootvetstviya
\begin{figure}
\begin{center}
  \includegraphics[width=70mm]{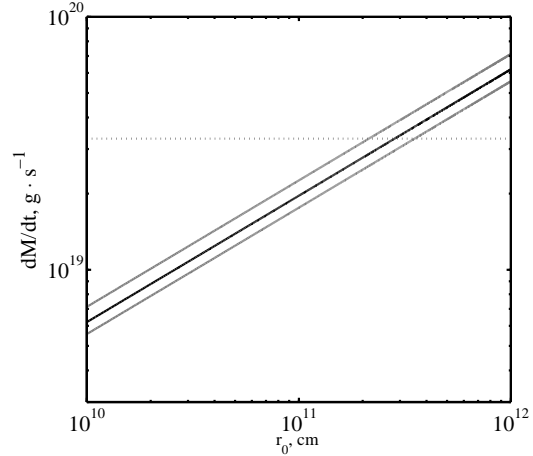}
\end{center}
 \caption{Determination of the blue jet visible base $r_0$ from the Fe\,XXV line flux
and the temperature derived (Table~1) for the observation in the precessional  
phase $\psi = 0$. To show confidence levels for $r_0$ (two solid grey lines) we used 
triple errors of the line flux. Horizontal dashed line corresponds to the mass loss rate
$\dot M_j = 3.3\times10^{19}$\,g/s ($L_{kin} = 10^{39}$\,erg/s).  
}
\label{F:r0}
\end{figure}

In our model the iron line flux (as it follows from (\ref{E:Jetspecta}, \ref{E:Norm})) 
may be written as:
\begin{equation}\label{E:FdM}
 F_l \propto J_l(\theta_0) \frac{\dot M_j^2}{\Omega_j r_0} \varphi(r_0, \dot M_j).
\end{equation}
Here $\varphi(r_0, \dot M_j)$ is a function depending on the gas cooling details and it does 
not strongly depend on $r_0$ and $\dot M_j$. At the same time the line ratio Fe\,XXV/Fe\,XXVI
depends on the gas temperature at the base $\theta_0$ and its derivative $d\theta / dr$, which 
is determined by (\ref{E:Teq}). Modelling of the iron line fluxes gives us information on
the temperature and visible jet base $r_0$.  

To determine the line fluxes in the observed spectra we model the continuum with power law,
and the lines as Gaussian blueshifted to $z=-0.105$ and $z=-0.095$ for precessional phases
$\psi=0$ and $\psi=0.8$ respectively. These blue shits were found in our fits, and they
are in approximative agreement with those expected from the precessional ephemeris. 
For better fits of these two lines with the continuum we discarded the spectral
parts below $< 7$~keV and $8.0-9.5$~keV, where there is a complex of emission lines   
Ni\,XXVII, Ni\,XXVIII, Fe\,XXVI\,L$\beta$ and Fe\,XXV\,K$\beta$. This spectral interval
also includes a broad absorption feature (an absorption edge), which was reported by 
\citet{kubota07}. In this simplest continuum model the line fluxes are well determined.
Table\,1 presents the fluxes of these two blueshifted lines and their ratios 
along with errors.

In Fig.\,\ref{F:ratio} we show the observed flux ratios Fe\,XXV\,/\,Fe\,XXVI along with 
the model ratio depending on the gas temperature at the jet base. We check how stable is the
temperature derived relating to the variations of $\dot M_j$ and $r_0$. Fig.\,~\ref{F:Tlimits}
shows the temperature as depending on $r_0$ at different mass loss rates $\dot M_j$ for the
observation taken in the precessional phase $\psi = 0$. These two parameters do not change the
base temperature notably. We find the same for the observations in the precessional phase $\psi = 0.8$.
One may expect this result because the line flux (\ref{E:FdM}) depends on $\dot M_j$ and 
$r_0$ in the same manner for both lines.
%%%%%%%%%%%%%%%%%%%%%%%%%%%%VST
Fig.\,~\ref{F:Tlimits} shows that to keep the same temperature  at the visible jet 
base $\theta_0 (r_0)$ (the same iron line ratio), with bigger $\dot M_j$ we have bigger
$r_0$ to have about the same gas density $n_0(r_0)$. The function $\varphi(r_0, \dot M_j)$
depends on the gas density, which is $n \propto \dot M_j / r_0^2$.

There is also some interplay between the estimates of $\theta_0 (r_0)$ and $r_0$ 
when we fix the mass loss rate in the jets. The emissivity of the Fe\,XXV line 
peaks at $\theta \sim 5 - 6$~keV, and the total line flux is $\propto n^2$.   
When one finds $\theta_0$ from the Fe\,XXV\,/\,Fe\,XXVI line ratio and determines 
the jet base $r_0$ from the Fe\,XXV line flux, one obtains the gas density
at the base $n_0(r_0)$. The higher the temperature $\theta_0$ derived with 
regard to the 5\,--\,6~keV value, the lower the gas density in the emissivity peak
%%%%%%%%%%%%%%%%%%%%%%%%%%%%VST
jet region, the less $r_0$ we obtain to match the observed line flux. This effect 
is seen in Fig.\,\ref{F:Tlimits}, and it is not strong.
%The goal of the 
%present paper is not to determine accurate jet parameters depending on the disc 
%orientation (the precessional phase), but rather to study the additional 
%components in the X-ray spectrum.

Thus, we may find the temperature at the visible jet raduis
$\theta_0(r_0)$ from the Fe\,XXV\,/\,Fe\,XXVI flux ratios, and then we may find the base 
$(r_0)$ from the observed flux of the blue shifted Fe\,XXV line and the derived 
temperature. The values of $\theta_0 (r_0)$ and $r_0$ found with the simple continuum 
model are presented in Table~1.

\begin{figure}
\centering
% \begin{center}
  \includegraphics[width=50mm]{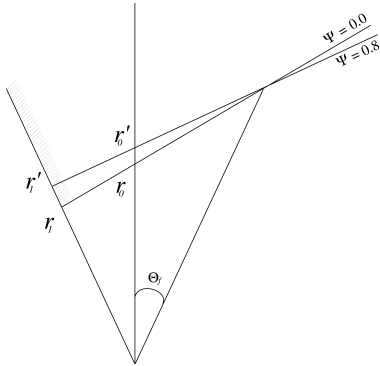}
 %\end{center}
 \caption{A sketch of the funnel and the blue jet visibility at two precessional phases
 %%%%%%%%%%%%%%%%%%%%%%%%%%%%VST
 considered. $r_0$ ($r_0^{\prime}$) is the distance between the base of the visible jet 
 and the top of the cone (the bottom of the funnel). $r_1$ ($r_1^{\prime}$) is the distance 
 between the base of the visible funnel wall and the top of the cone.}
 \label{F:geometry}
\end{figure}

Fig.\,\ref{F:r0} shows how the base $r_0$ is determined from the Fe\,XXV line flux. 
We used here the triple errors (Table~1) of the He-like blue jet line flux.   
The blue jet visible base we found is $r_0 \sim 3 \cdot 10^{11}$\,cm. 
%%%%%%%%%%%%%%%%%%%%%%%%%%%%VST
The equation (\ref{E:FdM}) provides us with useful and obvious scaling formula, 
$r_0 \propto \dot M_j^2 / \Omega_j$, and Fig.\,\ref{F:r0} confirms the scaling. 
With different jet parameters satisfying this scaling and with about the same 
gas temperature at the visible jet base, we obtain about the same thermal 
jet spectra.

Fig.\,\ref{F:geometry} shows a sketch of the jet and  
the supercritical accretion wind (funnel) geometry at two precessional phases
which we consider, $\psi \approx 0$ (orbit 707) and $\psi \approx 0.8$ (orbits
609, 610). They are marked in the figure as $r_0$ and $r^{\prime}_0$ respectively.
We adopted the funnel opening angle $\vartheta_f = 25^{\circ}$ \citep{Ohsuga05}
in the figure. Our estimates of $r_0$ and $r^{\prime}_0$ are the same within the 
errors (Table\,1), however the result depends on the fluxes of the 
Fe\,XXV line, i.~e. on the natural variability of SS\,433 jet activity. In the second 
observation at the $\psi \approx 0.8$ the iron line is brighter. 

The estimates of $\theta_0 (r_0)$ and $r_0$ obtained in this section were derived 
using the simplest continuum model. In this model we discarded the spectral 
part below $< 7$~keV and the region of $8.0-9.5$\,keV, which includes a 
broad absorption edge \citep{kubota07}. This absorption feature may distort the 
Fe\,XXV\,/\,Fe\,XXVI line ratio making it higher,
as a part of the Fe\,XXVI line profile may be inside of the broad edge. The 
edge profile is not established yet \citep{kubota07}, and we can not rule out that 
real temperature $\theta_0 (r_0)$ is somewhere higher than the one we derived here.   
Our estimate of $r_0$ is not affected by the absorption edge, because it is based
on the Fe\,XXV line. We use these estimates of $\theta_0 (r_0)$ and $r_0$ 
%%%%%%%%%%%%%%%%%%%%%%%%%%%%VST
(Table~1) as a first approximation. Below we find more accurate values 
of these parameters using a more complex model of the continuum, which describes 
the whole spectrum of SS\,433.

%%%%%%%%%%%%%%%%%%%%%%%%%%%%%%%%%%%%%%%%%%%%%%%%%%%%%%%%%%%%%%%%%%%%%%%%%%%
\section{Additional components in SS~433 spectra. Reflection model}
\label{reflection}
%%%%%%%%%%%%%%%%%%%%%%%%%%%%%%%%%%%%%%%%%%%%%%%%%%%%%%%%%%%%%%%%%%%%%%%%%%%
% nabl dok-va refl; ono d/b chastichno ionizovano; 
% ris s otrazheniem 5-7 i 7-10; opisanie; parametr ksi~300; tut N_H~1.5 10e22
% ono vliyaet tolko na soft
% podbiraem soft ris 5->6 bez dop zhestkogo komp (otrazheniya) i obsuzhdaem soft
% nahodim N_H i govorim o nalichii soft excessa
% perehodim k modelirovaniyu 4-h diap, t k padayuschii spectr m/b slozhnyh;
% no modeliruem s odinakovymi parametrami stenki

\begin{figure}
\center{\includegraphics[width=70mm, angle=0]{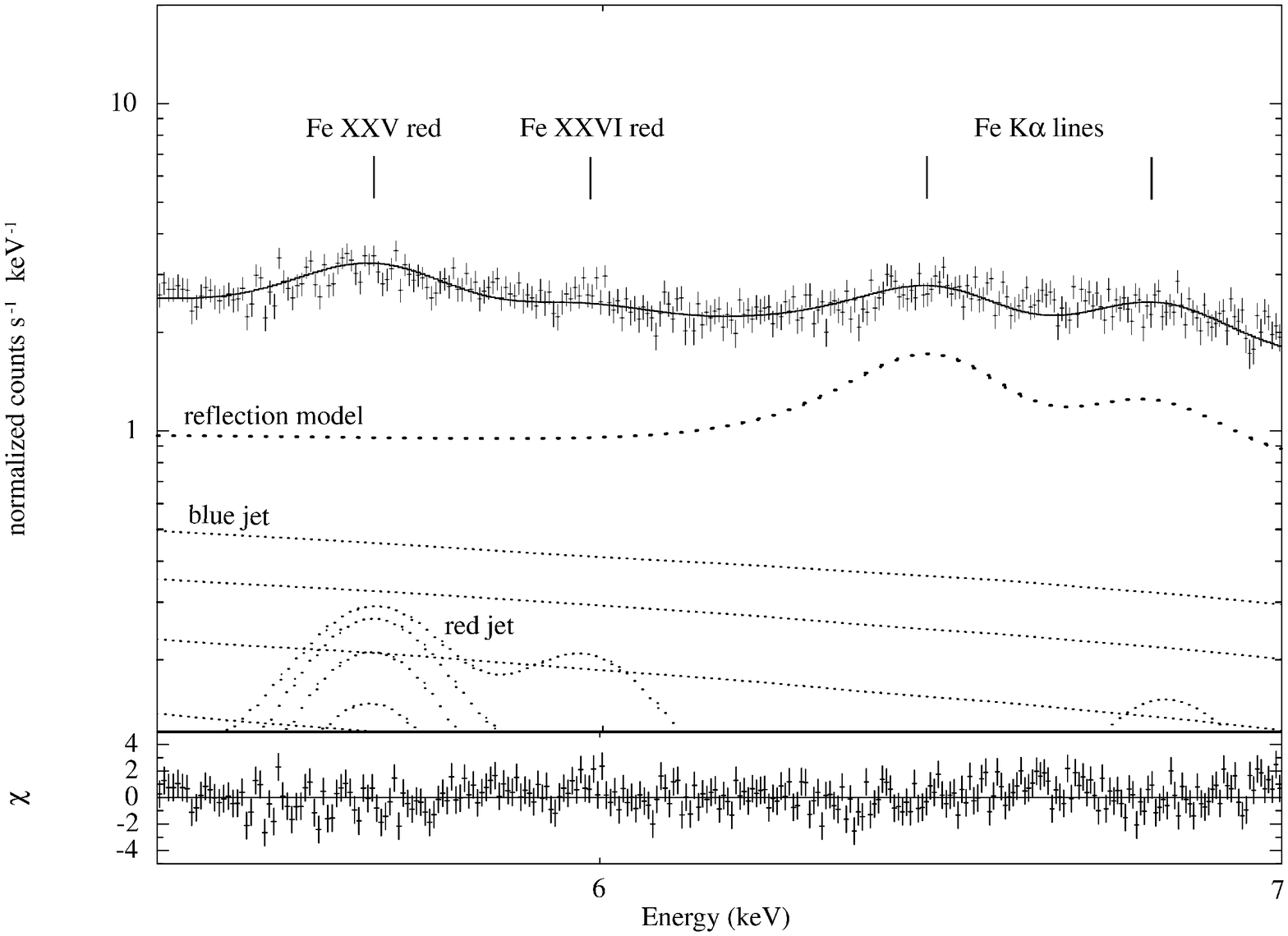}} \\
\vspace*{-0.5cm}
\hspace*{0.4cm}
\center{\includegraphics[width=72mm, angle=0]{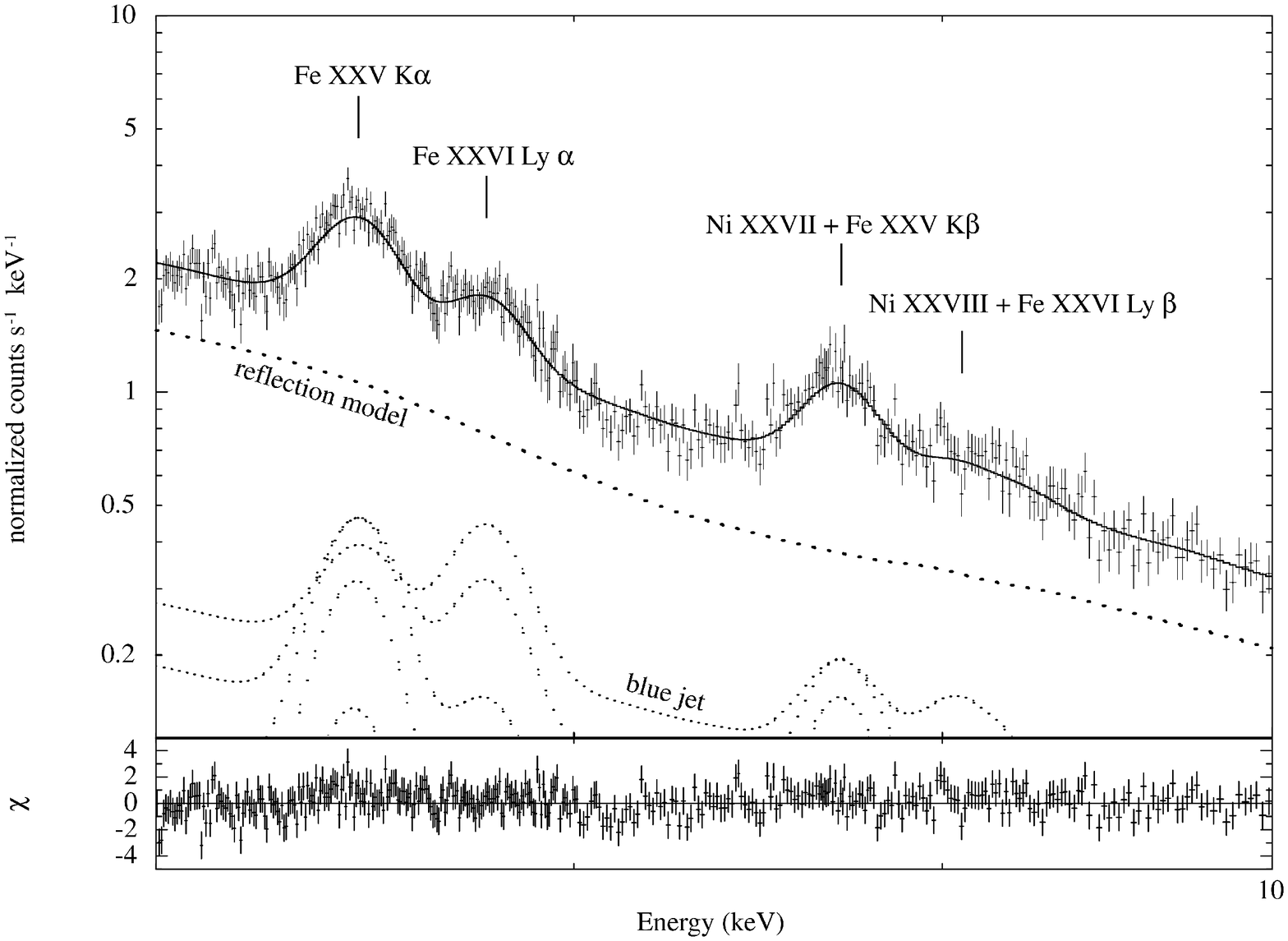}} \\
\caption{Observed spectrum of SS\,433 (orbit 707, $\psi \sim 0$) 
with the thermal jet model components and the additional reflection model 
in two spectral regions 5.0--7.0 (top) and 7.0--10.0~keV (bottom). The total
model spectrum is shown with solid line. The reflection model 
explains well the fluorescence iron line at 6.4~keV, the recombination iron 
K$\alpha$ line of Fe\,XXV at 6.7~keV (top) and the iron absorption 
edge \citep{kubota07} at $\sim 8-9$~keV.
}
\label{F:ReflIncreased}
\end{figure}

As it was noted in the Introduction, there is a problem of four orders of magnitude, 
missing in X-ray luminosity of SS\,433. One may expect to find the original luminosity 
of the object $L_x \sim 10^{40}$\,erg/s, instead, we observe $L_x \sim 10^{36}$\,erg/s. 
This may mean that direct radiation is blocked inside the hydrodynamical funnel,
where the jets are accelerated and collimated \citep{Fab04}. Taking into account
such an extreme original luminosity we may expect to observe a reflection of a part of the
radiation at the outer funnel walls (the wind). In Fig.\,\ref{F:geometry} we show
the probable reflecting area on the wall at $r > r_1$. 

There are strong observational indications that point to an existence of a reflection 
component in the X-ray spectrum of SS\,433. Firstly, the presence of a strong fluorescent line of 
quasi-neutral iron (6.4~keV), which can be a result of the reflection. Secondly, 
in the spectral region $\sim 8-9$~keV there is a broad absorption feature (a trough),
detected by \citet{kubota07}. The trough may be an iron absorption edge arising in 
reflection, its position corresponds not to the neutral plasma, but the some intermediate 
stages of iron ionization \citep{kubota07}. This leads to an assumption that
the reflecting matter in partly ionized. That is why we decide to fit the observed
spectrum of SS\,433 with the multi-temperature jet model plus reflection in the partly 
ionized material. 

For the reflection component we use the 
REFLION\footnote[1]{\texttt{http://heasarc.gsfc.nasa.gov/docs/xanadu/xspec/models/ \\
reflion.html}}
model of X-ray ionized reflection \citep{ross99,ross05}. 
The illuminating radiation is assumed to have a cut-off power-law spectrum 
$F_E = A E^{-\Gamma +1} \exp(-E/E_c)$, where the cut-off energy is fixed at $E_c=300$~keV.
It is assumed that the surface is irradiated with the X-rays so intense that its ionization state 
is determined by the ionization parameter,
\begin{equation}\label{E:ionpar}
\xi = \frac{4\pi F_{tot}}{n},   
\end{equation}
where $F_{tot} =\int F_E dE$ -- the total illuminating flux, $n$ -- the surface gas density.
The amplitude $A$ in the total illuminating flux is chosen to provide a desired value of 
the ionization parameter $\xi$. In the model the irradiated gas is Compton thick and has 
a constant density. The illuminating radiation flux has a power-law spectrum in the 
model, and the photon index $\Gamma$ is a model parameter.

Fig.\,\ref{F:ReflIncreased} shows the observed spectrum of SS\,433 
%(orbit 707, $\psi \sim 0$) 
with the thermal jet model and additional reflection model. It presents two regions 
5.0--7.0~keV and 7.0--10.0~keV fitted with the same model parameters. Both the reflection 
emission feature at 6.4~keV and the iron absorption edge at $\sim 8-9$~keV are reproduced
well in the reflection model. Fig.\,\ref{F:ReflIncreased} discovers one more an evidence
of the reflection in SS\,433 spectrum, it is an emission feature of iron Fe\,XXV\,K$\alpha$
at 6.7~keV at zero velocity. The REFLION model \citep{ross05} considers the K$\alpha$
recombination lines of Fe\,XXV--XXVI (near 6.7 and 7.0~keV) and the fluorescence lines 
of Fe\,VI--XVI (near 6.4~keV). The fluorescence features never been explained before in 
SS\,433 spectrum, this convinces us to use the reflection model as an additional (to the
thermal jet) component. In the reflection model fit we estimate the ionization parameter
at the reflecting surface of $\xi \sim 300$ and the illuminating radiation spectral index
of $\Gamma \sim 2$. The very good continuum fit 
%with the reflection around the jet iron lines 
allows us to estimate more accurately the iron line fluxes in the blue jet and respectively 
the visible jet base temperature, it is $\theta_0 (r_0) \sim 17$~keV. Below in this section   
we study these parameters in more details using the whole spectrum. 

The harder part of the spectrum $>5$~keV is not sensitive to the absorption value $N_H$, 
from the other hand, the reflection component we found weakens notably in softer energies, 
$<4$~keV. That is why to study the absorption value $N_H$, in a next step we analyse the whole 
spectrum of SS\,433 with the thermal jet component only.

\begin{figure} 
\begin{center}
\includegraphics[width=90mm,angle=0]{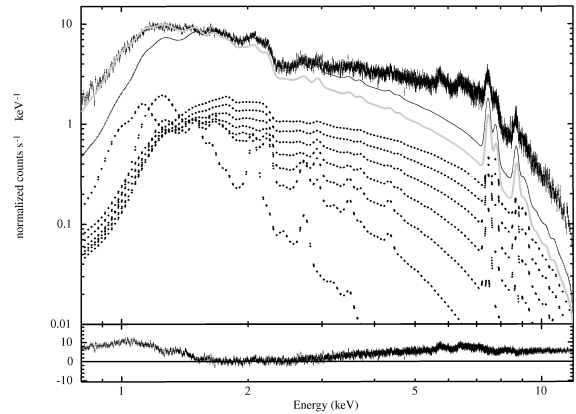}
\end{center}
\caption{Observed spectrum of SS\,433 (orbit 707, $\psi \sim 0$) with two thermal 
jet models. The first model ($\theta_0 (r_0)=17$~keV, $N_H=1.5\times10^{22}$\,cm$^{-2}$)
is shown with solid thin line, individual parts of the multi-temperature jet are shown
with dashed lines together with the model's residuals in bottom. The model produces well 
the blue jet lines in the region 7--9~keV. The second model ($\theta_0 (r_0)=14$~keV, 
$N_H=1.0\times10^{22}$\,cm$^{-2}$) is shown with grey thick line, this model is intended
to fit the lower energy range, however it fails to produce observed line fluxes of the
blue jet. The strong excess of the observed continuum radiation in the region 
$>2.5$~keV may be fited well with the reflection additional component (section 5).
}\label{F:whole}
\end{figure}

In Fig.\,\ref{F:whole} we show the observed spectrum of SS\,443 along with the 
thermal blue jet models.
%, whose parameters were found in the previous section. 
We only used in our figure the blue jet, because adding the red jet can not 
in effect solve the continuum problems both at higher energy and even at lower 
energies \citep{brinkmann05}. 
%%%
\citet{kotani96} have found that the receding red jet is not visible in low energy 
ranges (in the ranges $<4$~keV) in the precessional phases 
$\psi \sim 0$, when the accretion disc is the most open to the observer. 
They interpreted this fact via an existence of absorbing gas in the precessing plane 
perpendicular to the jets (the accretion disc plane), which only blocks the outer cooler parts
of the red jet. The hotter parts of the red jet located closer to the central 
machine, are visible, the extended material does not obscure them. The extended precessing 
disc-like outflow in SS\,433 has been discussed before \citep{Fabrika93}. 

There are two jet models in Fig.\,\ref{F:whole}, in both models we have adopted 
our standard jet parameters, $\dot M_j = 3.3 \cdot 10^{19}$\,g/s and $2\vartheta_j = 1.5^{\circ}$.
There are strong discrepancies between the observed and the model spectra.
Whereas the jet lines may be well fitted with one of the model (the Fe\,XXV and Fe\,XXVI 
lines of the approaching jet), the continuum divergence is dramatic.
This was noticed by \citet{brinkmann05}. They concluded that when the line spectrum 
is well reproduced by the thermal jet model, but the continuum is not. We tried to 
use our jet model with the parameters found by \citet{brinkmann05}, and confirmed 
their results.
When the hard energy region of the continuum 
can be fitted well by adding the reflection component, the soft energy range is not.

Producing the fits shown in Fig.\,\ref{F:whole}, we tried to match the spectrum best
possibly to illustrate the thermal jet model limitations. The first model 
fits well the iron line fluxes, however 
we observe a strong soft deficit. The neutral hydrogen column density 
$N_H$ in the fit is $1.5 \times 10^{22}$\,cm$^{-2}$ and the visible jet base 
temperature is $\theta_0 (r_0) = 17$~keV in the model. Note this column density 
is in very good agreement with the interstellar extinction in SS\,433 found in optical 
and UV observations \citep{Fab04}. To diminish the soft deficit one must take 
less temperature at the jet base. The best fit we can find in this way is  
shown in Fig.\,\ref{F:whole} as the second model ($N_H = 1.0 \times 10^{22}$\,cm$^{-2}$,
$\theta_0 (r_0) = 14$~keV). It has not such a big problem in the low energy region,
however the blue jet iron line fluxes can not be reproduced in this model, 
they are at least $\sim 1.5$ times less than the observed fluxes. Besides that,
the softest energy spectrum ($<1$~keV) is not descibed well even in the second model. 
We conclude that the thermal jet model can not describe simultaneously the line 
fluxes and the soft continuum.   
 
We have found that the thermal jet model produces about the same spectra when we 
keep the same value of $\dot M_j^2 / \Omega_j r_0 = const$. This scaling 
formula is valid in the first approximation only, $\varphi(r_0, \dot M_j)$ in 
(\ref{E:FdM}) does not strongly depend on $r_0$ and $\dot M_j$ (see previous section). 
For instance, when we compare the thermal jet model shown in Fig.\,\ref{F:whole} 
by thin solid line ($\dot M_j = 3.3 \cdot 10^{19}$\,g/s and $2\vartheta_j = 1.5^{\circ}$), 
with another model, where we take these two parameters twice as bigger 
(in a limit of small angles $\Omega_j \propto \vartheta_j^2$), we obtain the same 
model spectrum when we increase the value of $N_H$ by 17\,\%. Both the iron line 
fluxes and their equivalenth widths are the same in these two models. However the 
line profiles appear too broad in the second case, at least during this (orbit 707) 
observation the jets were notably narrower than $2\vartheta_j = 3.0^{\circ}$. 

%times higher than that found by 
%\citet{brinkmann05} ($(0.9-1)\times 10^{22}$\,cm$^{-2}$), probably because we 
%used only one approaching jet here. 
It is very important to note here that \citep{brinkmann05} have concluded that the 
thermal jet model in any combination can not produce the observed continuum. 
We confirm their conclusion here, and we may state the problem more concrete. 
When one reproduces well the continuum around the blue jet iron lines
($\sim 6-10$~keV), 
%%with the thermal jet and the reflection (as an additional hard component) models, 
one can measure the line fluxes without strong uncertainties. 
One can determine then the jet visible base temperatute and estimate 
the value of $\dot M_j^2 / \Omega_j r_0$ (\ref{E:FdM}) using the thermal jet model. 
We have found in this way that the thermal jet model can not produce the lower
energy continuum in SS\,433. However the strongest divergence is observed in higher
energy region, where about 50\,\% of the flux is formed not in the thermal jet
(Fig.\,\ref{F:ReflIncreased}). We have discussed above the clear evidences of the 
reflection in this spectral region.

\begin{figure*}
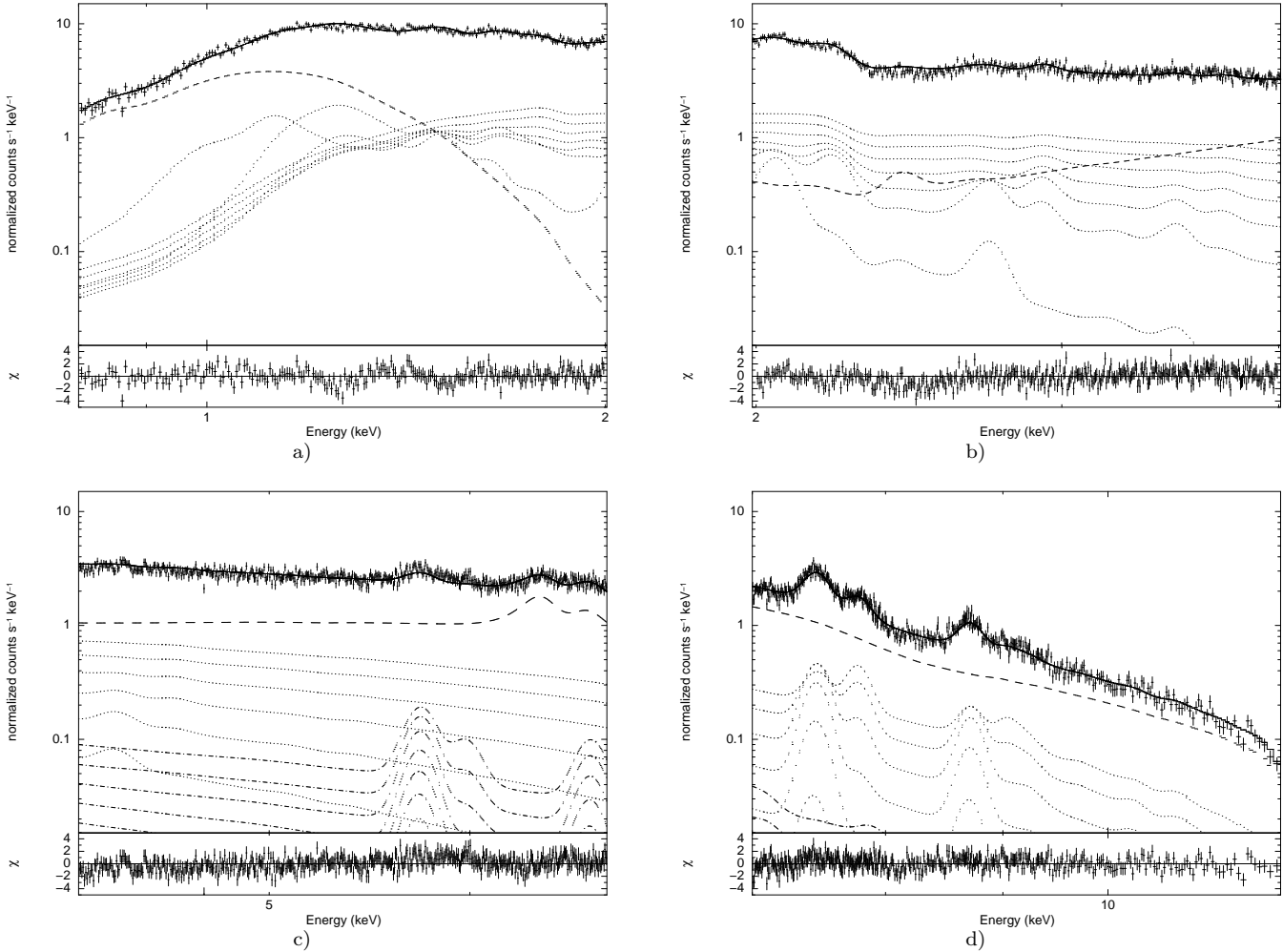

\begin{minipage}[h]{0.47\linewidth}
\center{\includegraphics[width=61mm, angle=-90]{pic/fig7a.ps}} a) \\
\end{minipage}
\hfill
\begin{minipage}[h]{0.47\linewidth}
\center{\includegraphics[width=61mm, angle=-90]{pic/fig7b.ps}} \\b)
\end{minipage}
\vfill
\begin{minipage}[h]{0.47\linewidth}
\center{\includegraphics[width=61mm, angle=-90]{pic/fig7c.ps}} c) \\
\end{minipage}
\hfill
\begin{minipage}[h]{0.47\linewidth}
\center{\includegraphics[width=61mm, angle=-90]{pic/fig7d.ps}} d) \\
\end{minipage}
\caption{
SS\,433 spectra (orbit 707, $\psi \sim 0$) with the model 'jets-plus-reflection' 
fit in four spectral ranges for case A, with zero red jet contribution in the 
ranges 0.8\,--\,4.0~keV (Table~\ref{T:approx}). The brightest jet components of 
the multi-temperature model are shown by
dotted lines for the blue jet and dash-dotted lines for the red jet, the total model
spectrum is shown by solid line. 
\textbf{a)} 0.8\,--\,2.0~keV range, where we added 
a black body spectrum to describe a soft excess observed at $\sim 1$~keV (dashed 
line). 
\textbf{b)} 2.0\,--\,4.0~keV, where the reflected spectrum (dashed line)
contribution weakens notably at lower energies.
%% ($\Gamma \la 1$). 
It is indicated by some excess in the observed spectrum at $\sim 2.2 - 2.6$~keV. 
\textbf{c)}
4.0\,--\,7.0~keV, where the reflection well describes both the continuum radiation, 
the fluorescent line of quasi-neutral iron and the recombination Fe\,XXV line
near 6.7~keV. Some excess at lower energies of the observed continuum is related to 
the change of the incident radiation spectral index (idem as in the previous range).  
\textbf{d)}~7\,--\,12~keV range is very well described by the reflection both in 
the continuum and in the broad absorption feature \citep{kubota07}.}
\label{F:ReflObs}
\end{figure*}

%% a mozhet byt chto pri obryve 300 kev my oshibemsya v otsenke padayuschego potoka

We know nothing about the incident flux and its spectrum
in SS\,433, so we divide the whole spectrum into four parts (0.8--2.0, 2.0--4.0, 4.0--7.0 
and 7.0--12.0~keV) and find the reflection model parameters independently in each part,
but with the same jet parameters and the same IS absorption $N_H$. 

The red jet spectrum has to notably contribute to the low energy region, however, 
it is not observed there \citep{kotani96}, but it is observed at higher energies 
(red iron lines). 
%%The low energy radiation is much more absorbed than high energy radiation.  
The lack of the red jet at low energies ($<4$~keV) may be bound not only to its specific
visibility, but with an underestimated extinction in the whole spectrum of SS\,433
as well.
From the XMM spectra available we can not realise clearly which effect is correct,
an obscuration of a cooler portion of the red jet or an underestimated extinction
in SS\,433. Therefore we produce two fits, first without the red jet cooler portion
in the range 0.8 -- 4.0~keV (case A),
%% probably more real)  
and second with the red jet in the whole spectrum with increased $N_H$ (case B). 
We find below that $N_H$ values derived in both cases are about the same and this
does not change our conclusions on the additional source of X-ray radiation in SS\,433. 

\begin{table*}
\begin{minipage}{140mm} \caption{
Results of the spectral fits in our four spectral ranges. The upper part corresponds
to the observation of orbit 707 ($\psi \sim 0.0$), the bottom part, -- to the orbit 609
($\psi \sim 0.8$). Case A is a model with a zero red jet contribution in softer 
(0.8\,--\,4.0~keV) ranges, case B with the red jet included. The jet parameters are 
$\theta_0 \approx 17$~keV, $r_{0,b} \approx 2\cdot10^{11}$\,cm, $r_{0,r} \approx 
7\cdot10^{11}$\,cm in the precessional phase $\psi \sim 0.0$, and $\theta_0 \approx 15$~keV
and the same visible jet base in $\psi \sim 0.8$. The lines present the 
ionization parameter (\ref{E:ionpar}) ($\xi$, erg/cm s), the photon index of a 
power law illuminating radiation $\Gamma$, the BB temperature 
in the soft X-ray range $\theta_{bb}$ (keV), the normalisation of the BB spectrum  
Norm$_{bb}=L_{39}/D^2_{10}$ ($L_{39}$ is luminosity in units $10^{39}$\,erg/s,
$D_{10}$ is a distance in units 10\,kpc), and column ISM density $N_H$
in units $10^{22}$\,cm$^{-2}$. $N_H$ is equal in each range as found
at lower energies.
}\label{T:approx}
\centering 
\begin{tabular}{cccccc}
\hline
 Range, keV                      &   & $7$-$12$        & $4$-$7$        &  $2$-$4$      & $0.8$-$2$ \\ 
\hline
%%$\Psi=0.8$  &   &    &   &      &  \\ 
\multirow{2}*{$\xi$}         & A & $300\pm 5$      &  $240\pm 30$    & $\approx 300$ & - \\
                             & B & $300\pm 4$      &  $250\pm 30$    & $\approx 300$ & - \\      
\multirow{2}*{$\Gamma$}      & A & $2.00\pm0.08 $  &  $1.56\pm 0.05$ & $\la 1$      & -  \\
                             & B & $2.00\pm 0.09 $ &  $1.57\pm 0.04$ & $\la 1$      & - \\
\multirow{2}*{$\theta_{bb}$} & A & -               & -               & -             & $0.107\pm 0.001$ \\
                             & B & -               & -               & -             & $0.100 \pm 0.001$ \\
\multirow{2}*{Norm$_{bb}$}   & A & -               & -               & -             & $0.032\pm 0.002$ \\
                             & B & -               & -               & -             & $0.125\pm 0.005$ \\
\multirow{2}*{$N_H$}         & A & -               & -               & -             & $1.54\pm 0.01$ \\
                             & B & -               & -               & -             & $1.85\pm 0.01$ \\
\multirow{2}*{$\chi_r^2$ (d.o.f)}&A& $1.05$ ($441$)&$1.28$ ($535$)   &$1.39$ ($396$) & $1.39$ ($242$) \\
                                 &B& $1.13$ ($441$)&$1.13$ ($555$)   &$1.55$ ($397$) & $1.58$ ($242$)\\
\hline
%%$\Psi=0.8$  &   &    &   &      &  \\   
\multirow{2}*{$\xi$}         & A & $340\pm 80$      &  $300\pm 90$    & $\approx 300$ &    - \\
                             & B & $310\pm 10$      &  $300\pm 90$    & $\approx 300$ &    - \\      
\multirow{2}*{$\Gamma$}      & A & $2.2\pm0.2$      &  $1.56\pm 0.08$ & $\la 1$      &    -  \\
                             & B & $2.0\pm 0.2 $    &  $1.57\pm 0.08$ & $\la 1$      & - \\
\multirow{2}*{$\theta_{bb}$} & A & -                & -               & -             & $0.093\pm 0.004$ \\
                             & B & -                & -               & -             &  $0.097 \pm 0.003$ \\
\multirow{2}*{Norm$_{bb}$} & A & -                & -               & -             & $0.026\pm 0.003$ \\
                             & B & -                & -               & -             & $0.068\pm 0.005$ \\
\multirow{2}*{$N_H$}         & A & -                & -               & -             & $1.28\pm 0.01$ \\
                             & B & -                & -               & -             & $1.58\pm 0.01$ \\
\multirow{2}*{$\chi_r^2$ (d.o.f)}&A& $1.36$ ($189$) &  $1.1$ ($332$)  &$1.01$ ($388$) & $1.55$ ($234$) \\
                                 &B&  $1.39$ ($190$)& $1.08$ ($322$)  & $1.05$ ($388$)& $1.05$ ($234$) \\
\hline
\end{tabular}
\end{minipage}
\end{table*}

We estimated the column density $N_H$ of the neutral IS absorbing material in the range 
0.8\,--\,2.0~keV. In this range a strong excess of radiation is observed 
(Fig.\,\ref{F:whole}), which does not practically depend on the thermal jet model
parameters. Hereat we added a BB component, which might be an indication of
the supercritical accretion disc funnel (a wind) radiation (see Section \ref{Disc}).
In Fig.\,\ref{F:ReflObs} we show the observed spectrum of SS\,433 along with the 
model fits 'jet-plus-reflection' in the four spectral ranges. Table\,\ref{T:approx} presents
the model parameters. The thermal jet model is identical in all ranges, yet the reflected 
component was fitted independently in each range (however there is no reflection in 
low energies).   

The blue jet parameters were found in the high energy range 7.0\,--\,12.0~keV
matching the model fluxes and flux ratios of the He- and H-like iron lines
to the observed values. We do not see any need to use non-solar abundances 
of the elements, but Ni is clearly overabundant. \citet{brinkmann05} found the Ni
abundance to be about 8 times solar. Our fits show that Ni has to be overabundant 
by a factor 10, since the Ni line
is inside the broad absorption feature (Fig.\,\ref{F:ReflObs}\,d). The red jet 
parameters were found in the range 4.0\,--\,7.0~keV, where the red 
iron lines are seen clearly (Fig.\,\ref{F:ReflObs}\,c). 

The jet parameters presented in Table\,\ref{T:approx} are more accurate than those
presented in Table\,1, where we used the simplest continuum model.
In the precessional phase $\psi \sim 0.0$ we find that the blue jet visible base  
is of $r_{0,b} \approx 2\cdot10^{11}$\,cm and the gas temperature in this place 
is $\theta_0 \approx 17$~keV. The red jet base is of $r_{0,r} \approx 
7\cdot10^{11}$\,cm, which is naturally explained by the obscuration of the red jet
by the accretion disc (the wind) body in this orientation of the disc. In the 
precessional phase $\psi \sim 0.8$, where one may expect an obscuration of a bigger
portion of the blue jet we find a lower temperature, $\theta_0 \approx 15$~keV. 
However, $r_{0,b}$ is about the same in both precessional phases. The spectra
taken at $\psi \sim 0.8$ are notably noisier than the one taken at $\psi \sim 0.0$. 

Finding accurate jet parameters is not the main purpose of the present paper. The results
depend on the natural variability in the jet activity, besides that the physical picture
may be more complex (possible gas clumping, heating effects due to kinetic 
energy dissipation or the funnel radiation). However, the derived jet visible 
base $r_{0,b} \sim 2 \cdot 10^{11}$\,cm is too small, while it has to be 
about $10^{12}$\,cm to satisfy the observed eclipses by the donor star. The inner hotter 
portions of the jet are partially eclipsed \citep{Kawai89,filippova06}. 
The jet's length is scaled (\ref{E:FdM}) as 
$r_0 \propto \dot M_j^2 / \Omega_j$, and one may scale the length to about 
the expected value of $\sim 10^{12}$\,cm in the thermal jet model which we use. 

The reflection model produces a very good fit in the high energy range 7.0\,--\,12.0~keV
($\chi^2 = 1.05$), it describes well both the whole continuum and the broad absorption 
feature detected by \citet{kubota07}. The feature is located between the blue shifted 
Fe\,XXVI and Ni\,XXVII lines and partly covers these lines (Fig.\,\ref{F:ReflObs}\,d). 
The parameters of the reflection are found, the ionization parameter of the irradiated
surface is $\xi\approx 300$ and the illuminating radiation index is $\Gamma \approx 2$. 
This means a flat ($E F_E$) spectrum of the incident radiation in this range.

In the range 4.0\,--\,7.0~keV we have an equally good agreement between the observed 
spectrum and the model. The reflection features at 6.2\,--\,6.8~keV are well 
produced. However, in this range there is a small continuum divergence at 
$\sim 4$~keV, probably due to the fact that the illuminating radiation index changes
with energy. We find the index $\Gamma \approx 1.6$ in the range, the 
ionization parameter is about the same as at higher energies, $\xi\approx 240$.

In the range 2.0\,--\,4.0~keV we observe the same effect is stronger. The model has 
a some deficit at 4~keV and an excess at 2~keV. An inclusion of the reflected component
is essential in the range, otherwise one can not produce the observed continuum at 
3\,--\,4~keV. The incident radiation has to have an index $\Gamma < 1$ in the range.
This is confirmed by a too strong emission feature at $\approx 2.5$~keV in the reflected
model. Similarly to the previous spectral range we may expect that the incident radiation
index is changed with energy, and here a steeper incident spectrum is expected.
The model we use is restricted by the value  $\Gamma = 1.0$. We find then at 
$\Gamma = 1.0$ the ionization parameter $\xi \approx 300$ (with a notable scatter). 

In the soft range 0.8\,--\,2.0~keV the reflected component is not detected. Even if
there is a reflection in the soft range, it has other parameters than those found
at higher energies. From the 7.0\,--\,12.0~keV band to the 2.0\,--\,4.0~keV band, 
we found a gradual decrease of the spectral index of the illuminating radiation, 
-- from $\Gamma \approx 2.0$ to $\Gamma \la 1$ at about the same ionization parameter, 
$\xi \sim 300$. However, an additional component related to the thermal jet one,
is indeed present in the soft energy range (Fig.\,\ref{F:whole} and 
Fig.\,\ref{F:ReflObs}\,a). Choosing the simplest way we add a BB component to 
describe the soft excess detected in the range.

The temperature of the BB component is $\theta_{bb} \approx 0.11$~keV and it is  
identical in  both our cases, A (without the red jet) and B (with the red jet). 
The BB component dominates at energies $\la 1$~keV and it is true again in both
cases. When we do not include the red jet in the fit, the BB luminosity is $\sim 4$
times smaller compared to the one with the red jet. However, this is rather a result 
of the increased $N_H$ in the latter case (B). The soft $\sim 1$~keV radiation is very 
sensitive to the $N_H$ value, therefore we can not estimate the real luminosity in the
soft excess from the XMM spectra. 

In both cases we find the IS extinction value to be about the same, it is 
(Table\,\ref{T:approx}) $N_H \approx 1.54 \times 10^{22}$\,cm$^{-2}$ in case A and 
$N_H \approx 1.85 \times 10^{22}$\,cm$^{-2}$ in case B. The interstellar extinction
in SS\,433, found in optical and UV observations \citep{Dolanetal97}, is
$A_V = 8.0 - 8.4$. This corresponds \citep{Predehl95} to $N_H=(1.4-1.5) \times 
10^{22}$\,cm$^{-2}$, which is in good agreement with the extinction found 
in X-rays. 

Spectral residuals in the soft energy range are bigger than the ones we 
have in other ranges. The most prominent negative feature which is not produced in 
the model fit is at 
1.2\,--\,1.3~keV (Fig.\,\ref{F:ReflObs}\,a). However, such a feature was predicted 
to be present in the spectra of ULXs, if they possess a supercritical accretion 
disc similar to the one in SS\,433, but observed nearly face-on. \citet{Fabr07} 
(see their Fig.\,2) have simulated spectra of the multi-colour funnel model (MCF), 
where they introduced an Lc edge of O\,VIII, blueshifted to the 
SS\,433 jet velocity. The absorption features arise inside the funnel. The model 
spectra being reduced with standard XSPEC models with IS absorption ($N_H \sim 
10^{22}$\,cm$^{-2}$) produce fake broad emission/absorption features. The strongest
one is an absorption feature at 1.2\,--\,1.3~keV. 

Our model fits to the spectra taken at the precessional phase $\psi \sim 0.8$
give us approximately the same results (Table\,\ref{T:approx}). We find a little lower
temperature and less luminosity of the soft BB component. In the sketch presented in 
Fig.\,\ref{F:geometry} the probable reflecting region is shown at $r > r_1$ at 
$\psi \sim 0$ and $r > r_1^{\prime}$ at $\psi \sim 0.8$ on the visible outer wall
of the funnel. If the soft excess is the visible funnel wall's own radiation  
(plus a possible reflection of a soft funnel radiation), one may expect less
luminosity of the wall due to the visibility conditions in the precessional 
phase $\psi \sim 0.8$.

%%%%%%%%%%%%%%%%%%%%%%%%%%%%%%%%%%%%%%%%%%%%%%%%%%%%%%%%%%%%%%%%%%%%%%%%%%%%%%%%%%%%%
\section{Discussion}
\label{Disc}
%%%%%%%%%%%%%%%%%%%%%%%%%%%%%%%%%%%%%%%%%%%%%%%%%%%%%%%%%%%%%%%%%%%%%%%%%%%%%%%%%%%%%

In this section we first discuss the structure of the supercritical accretion disc wind 
based on the wind approach by \citet{ShakSun73}, and estimate the wind and the 
jet funnel parameters for SS\,433. Next we discuss the reflected spectrum's origin and 
possible interpretations of the soft excess. 

%%%%%%%%%%%%%%%%%%%%%%%%%%%%%%%%%%%%%%%%%%%%%%%%%%%%%%%%%%%%%%%%%%%
\subsection{The expected parameters of the funnel in SS\,433}

\begin{figure} 
\includegraphics[width=80mm]{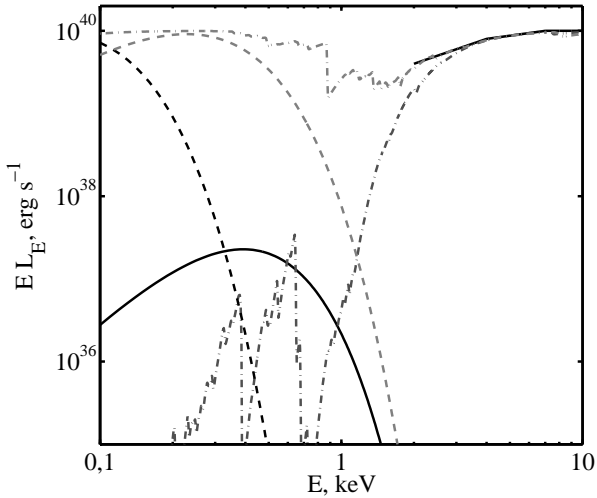}
\caption{Additional components in X-ray spectrum of SS\,433 (the thermal jet 
component is not shown). The illuminating incident radiation found in reflection
model is shown by solid lines (arbitrary scaled to $E L_E = 10^{40}$\,erg/s) in three
energy ranges: 2.0\,--\,4.0~keV ($\Gamma = 1.0$), 4.0\,--\,7.0~keV ($\Gamma = 1.6$)
and 7.0\,--\,12.0~keV ($\Gamma = 2.0$). The soft BB component 
($\theta_{bb} = 0.1$~keV, $L_x \sim 3 \cdot 10^{37}$\,erg/s) 
detected in the 0.8\,--\,2.0~keV energy range is shown 
by solid line. There are two originally the same flat spectra after absorption 
in ABSORI model with $N_H =10.0\times 10^{22}$\,cm$^{-2}$,
$\xi = 150$ (grey dash-dotted line) and  $N_H =6.0 \times 10^{22}$\,cm$^{-2}$,
$\xi = 3$ (bold dash-dotted line). There are two MCF (multi-colour funnel model) 
non-absorbed spectra with $r_{in} = r_1 = 1.74 \cdot 10^{11}$\,cm (bold dotted 
line) and $r_{in} = 5 \cdot 10^{10}$\,cm (grey dotted line).
}\label{F:add}
\end{figure}

The limiting Eddington luminosity is $L_E = 4\pi GMc/ \kappa = 1.5 \cdot 10^{39}\,m_{10}$\,erg/s,
corresponding limiting mass accretion rate is $\dot M_E = 48\pi GM/ck = 2 
\cdot 10^{19}\,m_{10}$~g/s, where the Thomson opacity for a gas with solar abundance 
is $\kappa = 0.35$~cm$^2$/g, and the black hole mass $m_{10}$ is measured in 10 solar mass units. 
With an accretion rate at the outer disc boundary $\dot M_a >> \dot M_E$, the disc becomes
supercritical at spherisation radius $r_{sp}$. In its outer parts $r \ga r_{sp}$ the disc is
subcritical, $\dot M(r) = \dot M_a$, below the spherisation radius $r < r_{sp}$ the accretion 
regime is locally critical, $\dot M(r) \sim (r/r_{sp}) \dot M_a$ and there appears a strong wind. 
The total mass loss in the wind below the spherisation radius is $\dot M_w(r) \sim \dot M_a (1-r/r_{sp})$.
The gas is blown away by the radiation pressure and the accretion rate is decreased linearly with 
radius. The outflow velocity is close to that of Keplerian, $v_w(r) \sim (GM/r)^{1/2}$, for 
the bulk of the wind it is $v_w \sim (GM/r_{sp})^{1/2}$. Following \citet{ShakSun73}, 
the spherisation radius is (also see \citet{Poutan07})
\begin{equation}\label{E:rad_sfer}
r_{sp} \approx  \frac{3 k \dot M_a}{8\pi c}. 
\end{equation}

This simple picture first drawn by \citet{ShakSun73} nevertheless explains the main features
that we observe in SS\,433 \citep{Fab04}. It is also confirmed in the main description in  
radiation hydrodynamic simulations \citep{Ohsuga05}. The simple model does not account for 
advection effects in the disc regions close to a black hole, however, this does not change 
the main picture. The $\dot M_w \sim \dot M_a$ 
%$\dot M_w(r) \sim \dot M_a (1-r/r_{sp})$ 
stays the same if $\dot M_a >> \dot M_E$, even the emerging spectrum of the supercritical 
disc does not change notably when advection is included \citep{Poutan07,Ohsuga09}. 

Adopting the mass exchange rate in SS\,433  $\dot M_w \sim \dot M_a \sim 10^{-4}\,M_{\sun}$/y
\citep{Shklov81,vandenHeuvel81,Fab04} we find the spherisation radius 
$r_{sp} \approx 10^{10} \, \dot m_{-4}$~cm and the wind velocity $v_w(r_{sp}) 
\sim 4000 \,(m_{10} / \dot m_{-4})^{1/2}$~km/s, where $\dot m_{-4}$ is measured 
in $10^{-4}\,M_{\sun}$/y units. The wind velocity is directly measured in SS\,433
\citep{Fab04}, it is $\sim 100$~km/s at the polar angle $90^{\circ}$ (disc edge-on) and 
increases sharply to $\sim 1500$~km/s at $60^{\circ}$ when the disc is 
the most open to observer.
%there are indications that at smaller polar angles the velocity is $v_w \sim 1500$~km/s.

Inside the spherisation radius the disc luminosity is locally Eddington, which gives
the total accretion luminosity \citep{ShakSun73} for SS\,433 $L_x \sim L_E (1+ 
\ln \dot M_a / \dot M_E) \sim 1.0 \cdot 10^{40}\,m_{10}$\,erg/s. Outside the funnel the 
radiation is thermalised ($L_{bol} \sim L_x$),
the wind's photospheric radius and its temperature in the photosphere are estimated as
\begin{equation} \label{eq:RTphot_w}
%\mbox{ }
\begin{aligned}
&r_{ph}= \kappa \dot M_{w} /4 \pi \cos \vartheta_f v_w \sim 1 \cdot 10^{12} ~ \rmn{cm}, \\  
&T_{ph}=(L_{bol}/ 4 \pi \sigma \cos \vartheta_f r_{ph}^2)^{1/4} \sim
          6 \cdot 10^4 ~ \rmn{K}.
\end{aligned}
\end{equation}
Both the size of the hot body and its temperature are quite close to those we observe 
in SS\,433. Inside the funnel where the jets are formed, the photospheric radius
(the funnel bottom) is  $r_{ph, j} = \kappa \dot M_j / \Omega_f v_j \sim 2.5 \cdot 10^9$\,cm,
where $\Omega_f = 2 \pi (1- \cos \vartheta_f)$ is the funnel's solid angle and
$\vartheta_f = 25^{\circ}$. The SS\,433 jet funnel is not transparent down to the
black hole. Below $r_{ph, j}$ the funnel walls are probably not seen directly
by a face-on observer.

A multicolour funnel model (MCF) has been proposed for SS\,433 \citep{Fab_etal06,Fabr07} to
estimate the emerging X-ray spectrum. They considered the opaque disc wind and estimated
the gas cooling and its temperature inside the wind.
%% (both with and without the funnel).
In the inner parts of the wind the radiation pressure dominates and $T(r) \propto r^{-1/2}$,
in the outer parts the gas pressure dominates and $T(r) \propto r^{-1}$. The supercritical
disc simulations \citep{Ohsuga05} show that the $T(r) \propto r^{-1/2}$ is valid
from $r \sim 20 \,r_g$ (below which the advective energy transport is important) to
$r \sim 200 \,r_g \sim 10^9$~cm, approximately the outer boundary of the computational 
domain.

One may estimate the radiation trapping radius in the wind outflow
comparing the dynamical time-scale $t_{mov} \sim r/v(r)$
and the photon diffusion time-scale $t_{esc} \sim n(r) r^2 \sigma_T /c$,
where $n(r) = \dot M_{w}/4 \pi \cos\vartheta_f m_P r^2 v(r)$, $\sigma_T$ is Thomson
cross-section, and $m_P$ is proton mass. The critical trapping radius in
the wind (where $t_{esc}=t_{mov}$) is $r_{tr} \sim (2/\cos\vartheta_f) r_{sp}
\sim 2 \,r_{sp}$. One may expect \citep{Fabr07,Poutan07} that radiation is trapped
within the wind up to $r \sim 2 \,r_{sp}$. In this region the radiation transfer is local.
In outer regions $r >> 2 \,r_{sp}$ the radiation transfer inside the wind is global
and radiation cools the wind ($T(r) \propto r^{-1}$).

This simple approach allows us to estimate (see below) the soft excess in the spectrum observed
in the low energy range (Fig.\,\ref{F:ReflObs}\,a). Adopting the wind temperature of
$T_{ph} \sim 60000$~K at the outer photosphere one may expect ($T(r) \propto r^{-1}$)
the temperature of $T \sim 0.5$~keV at the level of spherisation radius.

Photons entering the funnel from the funnel's walls have multiple scatterings on
the walls before they may finally escape the funnel. The probability to escape
from the infinite funnel after one scattering is $P = 1 - \cos \vartheta_f$. The probability
for a photon to remain in the funnel after N scatterings is $(1-P)^N$. If we consider
that 90\,\% of photons escaped the funnel ($\cos^N \vartheta_f = 0.1$), we find
the average number of scatterings $N(90\,\%) \sim 23$,
%A number of scatterings in an infinite funnel is $N \sim 1/(1-\cos \vartheta_f) \sim 10$
when $\vartheta_f = 25^{\circ}$. The number of scatterings in SS\,433 funnel
($r_{ph}/r_{ph, j} \sim 1000$) is not significantly smaller than that in the infinite funnel.

At the wall temperature of less than a few keV, as we may expect, the absorption is more
probable than electron scattering. This means that the photons are re-radiated back to the
funnel with the local wall temperature $T(r)$. The multicolour funnel has to have an
integral spectrum \citep{Fab_etal06,Fabr07} with numerous edges, where the strongest
edges are from the most abundant elements like iron and oxygen. The underlying spectrum
with the edges formed at about zero velocity is absolutely necessary to produce the observed
stable jet outflow velocity $v_j \approx 0.26 c$ in the line-locking mechanism
\citep{shapiro86,Fab04}.

%%%%%%%%%%%%%%%%%%%%%%%%%%%%%%%%%%%%%%%%%%%%%%%%%%%%%%%%%%%%%%%%%%
\subsection{The reflected and black body component}

Both the reflected component which we detect in three energy ranges (Fig.\,\ref{F:ReflObs},
Table~\ref{T:approx}), and the BB component in the soft energy range 0.8\,--\,2.0~keV,
may be formed at the outer visible part of the wall shown in Fig.~\ref{F:geometry}
at $r \ga r_1$. At the precessional phase $\psi \approx 0$ we find
$r_1 \approx 0.87\,r_0 \approx 1.74 \cdot 10^{11}$\,cm (Table~\ref{T:approx}). The
ionization parameter (\ref{E:ionpar}) found at the reflecting surface is $\xi \sim 300$
and it is approximately the same in all the three energy ranges. As we discussed above,
the multiple scattering in the funnel is expected, so we may consider this value of
$\xi$ as effective or averaged over all putative scatterings in the funnel. A study of
the additional spectral components is out of the scope of this paper, we only make
rough geometrical estimates here.

We estimate an angle $\beta$ at which the reflecting medium is exposed to the
incident radiation to produce the derived ionization parameter (\ref{E:ionpar}),
$\beta \sim \xi n r^2/L_x$. Using $n(r)r^2 = \dot M_{w}/4 \pi \cos\vartheta_f m_P v_w$,
we find $\beta \sim \xi \dot M_{w}/4 \pi \cos\vartheta_f m_P v_w L_x \sim
0.04\, \dot m_{-4} \,v_{2000}^{-1} \,L_{40}^{-1}$ for $\xi =300$, where the wind 
velocity is measured in units of 2000~km/s and the funnel luminosity in $10^{40}$\,erg/s
units. This corresponds to the funnel depth $\beta r_1 \sim 7 \cdot 10^9$~cm,
which is close to the spherisation radius.

The reflected component luminosity found in the energy range of 7\,--\,12~keV is
$L_{refl} \sim 3.3 \cdot 10^{35} \, k$\,erg/s (at a distance of 5.5~kpc), where
the $k$ factor is ratio of the total cone surface at $r > r_1$ (Fig.\,\ref{F:geometry})
to the visible cone surface. At the precessional phase $\psi \approx 0$, the ratio is
$k \approx 3.2$, and the luminosity at $r > r_1$ is  $L_{refl} \sim 1.0 \cdot 10^{36}$\,erg/s.
In the range of 7\,--\,12~keV the reflection albedo is mainly geometrical and we might expect
the angle $\beta \sim F_{refl}/F_{inc}$, where $F_{inc}$ is the incident flux in
this energy range, if the reflecting surface is directly exposed to the hard radiation
from its origin. Assuming that the illuminating spectrum is flat, with total 
luminosity of $\sim 10^{40}$\,erg/s in the spectral range of 0.1~--20~keV, we find that
its luminosity in the 7\,--\,12~keV range is $F_{inc} \sim 10^{39}$\,erg/s 
and the angle $\beta \sim F_{refl}/F_{inc} \sim 0.001$. This gives us an estimate
of the funnel depth, where the illuminating radiation is formed, 
$\beta r_1 \sim 2 \cdot 10^8$~cm.

The funnel photosphere radius, $r_{ph,j} \sim 2.5 \cdot 10^9$\,cm, derived from 
the observed values ($\dot M_j, v_j$), is notably greater than the last estimate of
the radius where the hard illuminating radiation is formed. Indeed, the reflecting surface 
may not directly see the deepest funnel parts. There have to be multiple scatterings 
in the funnel where the harder ($> 7$~keV) radiation may survive absorption. 

In Fig.\,\ref{F:add} we show the illuminating spectrum 
found above in the 4~--~12~keV energy range, arbitrarily scaled to the luminosity
of $10^{40}$\,erg/s. We show the spectral slope in the 2~--~4~keV 
range $\Gamma = 1.0$, but it was noted above that it has to be steeper, 
$\Gamma < 1$. The figure also shows two originally flat absorbed spectra.   
These two absorbed spectra illustrate two limiting cases between which one may
expect to find true picture. Even very thick ($N_H \sim 10^{23}$\,cm$^{-2}$)
medium of higher ionization ($\xi = 150$) can not depress the radiation spectrum 
below 3~--~4~keV. Lower ionization medium ($\xi = 3$) does absorb the radiation
effectively. 

Therefore we expect an existence of an absorbing (reflecting) 
medium inside the funnel, which produces such a curved
%% from its soft side 
spectrum of radiation, illuminating the reflected surface. 
Note that having He- and H-like absorption 
edges at about zero velocity is a mandatory property of the 
funnel spectrum to produce the observed stable jet velocity 
($v_j \approx 0.26$\,c) due to the line-locking mechanism 
\citep{shapiro86,Fab04}.  

Inspecting the results of the supercritical accretion disc simulations by 
\citet{Ohsuga05} we find that the walls at the level of $\sim 10^8-10^9$~cm consist 
of a low velocity gas with the temperature of $T \sim 2 \cdot 10^6$~K and the density 
of $n \sim (1-3) \cdot 10^{18}$\,cm$^{-3}$. The walls are the source of strong 
wind, they supply the funnel with the gas. The funnel walls' photosphere with 
the temperature from $T \sim 10^7$~K at $r \sim 10^8$\,cm to 
$T \sim 10^6$~K at $r \sim 10^9$\,cm, may be the source of the edged radiation spectrum. 

The nearly flat spectrum of the funnel (Fig.\,\ref{F:ReflObs},\,\ref{F:add}), 
which is expected to illuminate the reflecting surface, extends at least from 
4 to 12~keV. The supercritical accretion disc spectrum is expected to be flat
\citep{Poutan07}, however, it is supposed to go up to a few keV only. The Comptonization 
may extend and flatten the spectrum to higher energies. At an accretion rate 
of $\sim 10\,\%$ of the Eddington rate, the local emitted spectrum in the inner region 
of the disc is approximated by a diluted BB with the spectral hardening 
factor $f \sim 1.7$ \citep{shimura95}. For the accretion rates close to the Eddington 
limit, $f$ increases and may amount to 10. This may extend the disc spectrum 
to higher energies.

Inside the disc spherisation radius, the disc wind supplies the funnel with gas
starting from very deep regions \citep{ShakSun73,Eggumetal88,Okudaetal05,Ohsuga05},
this gas is directly accelerated by the funnel radiation. The interaction of the gas
accelerated to the velocities of $\sim 0.2 - 0.3$\,c with new portions of the walls' wind 
creates the conditions quite suitable for Comptonization. These conditions might be 
about the same throughout the whole funnel length where the jet is formed, as 
the high-speed outflow has about the same velocity. 

In the INTEGRAL observations a hard X-ray emission (20~--~100~keV)
with the total hard luminosity of $\sim 10^{35}$\,erg/s has been detected 
\citep{Cheretal05}, which may be well explained \citep{Krivoshhev09} in terms of 
Comptonized thermal plasma surrounding the visible jet base. 

Despite the fact that the jet formation and collimation mechanism is not clear yet,
one may expect the Comptonization is significant inside the funnel. The observed 
spectrum, taken when the disc is the most open to the observer \citep{Krivoshhev09},
is nearly flat in the energy range of 10\,--\,20~keV. This spectrum may be a continuation 
of the flat 7\,--\,12~keV spectrum found here in the reflecting model.
  
The soft excess detected in the SS\,433 spectra and introduced here as a BB
component with the temperature of $\theta_{bb} \approx 0.1$~keV (Table\,\ref{T:approx}, 
Fig.\,\ref{F:add}), may appear to be the direct radiation of the visible funnel wall at 
$r>r_1$ (Fig.\,\ref{F:geometry}). When we take into account the factor 
$k \approx 3.2$, which is a ratio of the total cone surface at $r > r_1$ 
to the visible cone surface, we find the soft BB component luminosity 
$L_x \sim 3 \cdot 10^{37}$\,erg/s.
 
The soft BB component is shown in Fig.\,\ref{F:add}. Note that the luminosity 
of the soft component strongly depends on the adopted column density $N_H$. 
It is essential here that the soft component is necessary 
\citep[][and herein]{brinkmann05} in order to understand the SS\,433 
spectra within the framework of the thermal jet model. 

In Fig.\,\ref{F:add} we show two versions of the funnel's non-absorbed 
spectrum in the MCF model \citep{Fab_etal06,Fabr07},
at $r> 1.74 \cdot 10^{11}$\,cm (the value of $r_1$ derived above) and 
$r> 5 \cdot 10^{10}$\,cm. Deriving these spectra, we adopted  
the wind's photosphere radius and the temperature of $r_{ph} = 10^{12}$\,cm, 
$T_{ph} = 50000$~K respectively (\ref{eq:RTphot_w}), and 
$T(r) \propto r^{-1}$ as it was discussed above. This illustrates that the 
MCF spectra might be responsible for the soft excess observed in SS\,433.
Nevertheless, one can say nothing else, exept that the additional soft 
component is present in SS\,433 spectrum. 

The soft excess is observed in the spectra of ULXs \citep{Stobbart06,Roberts07,Poutan07},
their temperature is $\theta \sim 0.1$~keV and it contributes to the total ULX 
luminosity of $\sim 10\,\%$. The soft component luminosity in ULXs is about 
$10^{38}-10^{39}$\,erg/s. If the ULXs are nearly face-on SS\,433 stars,
their intrinsic soft component luminosity has to be some less because of the
geometrical beaming of their supercritical discs. However, the soft X-ray luminosity
is model dependent, it strongly depends on the adopted absorption value.

%%%%%%%%%%%%%%%%%%%%%%%%%%%%%%%%%%%%%%%%%%%%%%%%%%%%%%%%%%%%%%%%%%%%%%%%%%%%%%%%%%%%%
\section{Conclusion}
%%%%%%%%%%%%%%%%%%%%%%%%%%%%%%%%%%%%%%%%%%%%%%%%%%%%%%%%%%%%%%%%%%%%%%%%%%%%%%%%%%%%%

The goal of the present paper was to find indications of the supercritical accretion disc 
funnel in SS\,433 X-ray radiation. The bolometric luminosity of the disc in SS\,433 is 
$\sim 10^{40}$\,erg/s and all the luminosity must be initially released in X-rays. 
This is confirmed by the well-established kinetic luminosity of SS\,433 jets, which 
is $\sim 10^{39}$\,erg/s, 
and the fact that the jets are formed in the deepest places of the disc funnel 
close to the black hole. The observed X-ray luminosity of SS\,433 is $\sim 10^{36}$\,erg/s
and it is the radiation of the cooling X-ray jets. Both the orientation of SS\,433 and the 
visibility conditions do not allow us to see any deep areas of the funnel. However,
the strong mass loss of the accretion disc gives us a hope for detecting some 
indications of the funnel radiation.

We have analysed the XXM spectra of SS\,433 with a well-known standard model of the
adiabatically cooling X-ray jets, taking into account cooling by radiation.
We confirm that the thermal jet model reproduces the emission line fluxes quite well.
When the disc is most open to the observer, the visible blue jet base is
$r_{0} \approx 2\cdot10^{11}$\,cm, the gas temperature at $r_{0}$ is 
$\theta_0 \approx 17$~keV. The IS gas column density is 
$N_H \sim 1.5 \cdot 10^{22}$\,cm$^{-2}$, which is in good agreement with the IS
extinction found in optical and UV observations. We confirm the previous finding 
\citep{kotani96,brinkmann05} that the red jet is probably not seen (blocked)
in the soft energy range of 0.8~--2.0~keV. However, both the model with blocked
the red jet portions, and the model with the whole red jet visible (but with some 
higher $N_H$) give the same result, that the thermal jet model alone can not 
explain the soft continuum. We also confirm that Nickel is highly 
overabundant in the jets \citep{kotani96,brinkmann05}. 

%%We do not try to determine the jet kinetic luminosity (the mass loss rate) in this 
%%analysis, as it may be not accurate due to the natural jet variability. Instead, 
We have adopted the luminosity $L_k = 10^{39}$\,erg/s and the jet opening angle 
$2\vartheta_j = 1.5^{\circ}$ as well established and known. However, the
derived jet visible base $r_{0}$ is too small, it has to be 
$\sim 10^{12}$\,cm to satisfy the observed eclipses by the donor star. Using the
scaling formula $r_0 \propto \dot M_j^2 / \Omega_j$ one may find that during these
XMM observations the jet kinetic luminosity was $L_k \sim 2 \cdot 10^{39}$\,erg/s.
This scaling should not change the thermal jet spectrum.

We find that the thermal jet model alone can not reproduce the continuum radiation
in the XMM spectral range. Therefore we use the thermal jet model together with 
the REFLION ionized reflection model and find quite a good representation of the
spectra. Introducing the reflection not only explains the continuum curvature
at the energies of $> 3-4$~kev and the fluorescent line of quasi-neutral iron, 
it reproduces the broad absorption feature at $\sim 8$~keV recently detected
by \citet{kubota07}. We independently study three energy ranges (2\,--\,4,
4\,--\,7 and 7\,--\,12~keV) with the same thermal jet model to find
the reflection parameters. We find that the ionization parameter (\ref{E:ionpar})
is about the same in these ranges, $\xi \sim 300$, which indicates a highly
ionized reflection surface. The illuminating radiation photon index changes 
from flat, $\Gamma \approx 2$, in the 7\,--\,12~keV range to $\Gamma \approx 1.6$
in the range 4\,--\,7~keV and to $\Gamma \la 1$ in the range 2\,--\,4~keV.

We conclude that the additional reflected spectrum is an indication of 
the funnel radiation. The illuminating radiation spectrum is flat in the range 
of 7\,--\,12~keV. With multiple scatterings in the funnel the hard radiation
may survive absorption. The reflected luminosity in this 7\,--\,12~keV
spectral range is $L_{refl} \sim 10^{36}$\,erg/s. The softer ($2-7$~keV) part 
of the illuminating spectrum (Fig.\,\ref{F:add}) carries a trace of 
absorption. This is assumed to be due to the multiple scatterings in 
the funnel. It is important to note that an existence of He- and H-like absorption edges 
at about zero velocity is a mandatory property of the funnel spectrum to be able to
produce the observed jet velocity due to the line-locking mechanism. 

The supercritical accretion disc spectrum is expected to be flat, however, it 
extends up to a few keV only. Comptonization may extend and flatten the spectrum 
to higher energies. The interaction of the high velocity gas moving along the jet axis 
with the walls' wind makes the conditions for the Comptonization throughout
the whole funnel length consistent. Direct evidences of the Comptonized radiation have
recently been found in the INTEGRAL observations \citep{Cheretal05}.

We have not found any evidences of the reflection in the soft 0.8\,--\,2.0~keV
energy range. The soft excess is detected in the SS\,433 spectra. We suppose that
in the soft X-rays we observe direct radiation of the visible funnel wall.
We represented this component as a black body radiation with a temperature of
$\theta_{bb} \approx 0.1$~keV and luminosity of $L_{BB} \sim 3 \cdot 10^{37}$\,erg/s.
The soft excess may be also fitted with a multicolour funnel (MCF) model.
Both the soft X-ray luminosity and its derived spectrum strongly depend on
the column density $N_H$. The soft component is necessary to add in order to 
understand the SS\,433 spectra within the framework of the thermal jet model.

The soft excess is observed in the spectra of ULXs with the soft component
temperature of $\theta \sim 0.1$~keV, which is identical with that found 
in SS\,433. If the ULXs or some of them are nearly face-on versions of 
SS\,433, one may adjust their soft X-ray components to the outer funnel 
walls radiation \citep{Poutan07}.

%%%%%%%%%%%%%%%%%%%%%%%%%%%%%%%%%%%%%%%%%%%%%%%%%%%%%%%%%%%%%%%%%%%%%%%%%%%%%%%%%%%%%
\section{Acknowledgements}
%%%%%%%%%%%%%%%%%%%%%%%%%%%%%%%%%%%%%%%%%%%%%%%%%%%%%%%%%%%%%%%%%%%%%%%%%%%%%%%%%%%%%

The authors thank Victor Doroshenko for useful comments and discussions, and Anna Zyazeva
for the correction of this manuscript. The research was supported by RFBR grant 07-02-00909.

\label{lastpage}
\end{document}